\documentclass[11pt]{article}

\newlength{\abstractwidth}
\usepackage{hyperref}
\usepackage{color}
\usepackage{graphicx}
\usepackage{tikz}
\usepackage{mathtools}
\usepackage{float}

\usepackage{hyperref}

\usepackage[most]{tcolorbox}

\tcbset{colback=white, colframe=red!50!black, 
        highlight math style= {enhanced, 
            colframe=red,colback=red!10!white,boxsep=0pt}
        }

\usepackage{amsmath}
\usepackage{amssymb}
\usepackage{latexsym}
 \usepackage{setspace}
 
 \usepackage{caption}
\usepackage{subcaption}
\usepackage{bbm}

\numberwithin{equation}{section}

\renewcommand{\thefootnote}{\fnsymbol{footnote}}
\renewcommand{\thanks}[1]{\footnote{#1}}
\newcommand{\starttext}{
\setcounter{footnote}{0}
\renewcommand{\thefootnote}{\arabic{footnote}}}
\newcommand{\bea}{\begin{eqnarray}}
\newcommand{\eea}{\end{eqnarray}}
\newcommand{\be}{\begin{eqnarray}}
\newcommand{\ee}{\end{eqnarray}}

\def\ie{\begin{equation}\begin{aligned}}
\def\fe{\end{aligned}\end{equation}}
\def\half{{\scriptstyle \frac 12}}

\def\threeh{{\scriptstyle \frac 32}}

\def\ie{\begin{equation}\begin{aligned}}
\def\fe{\end{aligned}\end{equation}}

\def\cC{{\cal C}}

\def\cN{{\cal N}}
\def\cO{{\cal O}}

\def\cQ{{\cal Q}}

\def\cS{{\cal S}}
\def\cT{{\cal T}}

\def\Z{{\mathbb Z}}

\def\RR{{\mathbb R}}

\def\Gnek{\mathcal{G}}

\def\nn{\nonumber}

\def\C{\cC}

\DeclareMathSymbol{\shortminus}{\mathbin}{AMSa}{"39}

\begin{document}

\starttext

\setcounter{footnote}{0}


\vskip 0.3in

\begin{center}

{\bf {\LARGE \sc 
Exceptionally simple integrated correlators\\
\vspace{0.25cm}

in $\mathcal{N}=4$ supersymmetric Yang--Mills theory}}

\vspace{6mm}
\normalsize
{\large  Daniele Dorigoni$^{F_4}$ and Paolo Vallarino$^{G_2}$}

\vspace{10mm}
$(F_4)\,\,${\it  Centre for Particle Theory \& Department of Mathematical Sciences\\
Durham University, Lower Mountjoy, Stockton Road, Durham DH1 3LE, UK}\\
\vspace{0.3cm}
$(G_2)\,\,${\it Universit\`a di Torino, Dipartimento di Fisica and I.N.F.N. \\
- sezione di Torino Via P. Giuria 1, I-10125 Torino, Italy }

\vskip 0.5in

\begin{abstract}
\vskip 0.1in

Supersymmetric localisation has led to several modern developments in the study of integrated correlators in $\mathcal{N}=4$ supersymmetric Yang--Mills (SYM) theory.
In particular, exact results have been derived for certain integrated four-point functions of superconformal primary operators in the stress tensor multiplet which are valid for all classical gauge groups, $SU(N)$, $SO(N)$, and $USp(2N)$, and for all values of the complex coupling, $\tau=\theta/(2\pi) + 4\pi i  /g^2_{_{YM}}$.
In this work we extend this analysis and provide a unified two-dimensional lattice sum representation valid for all simple gauge groups, in particular for the exceptional series $E_r$ (with $r=6,7,8$), $F_4$ and $G_2$. These expressions are manifestly covariant under Goddard-Nuyts-Olive duality which for the cases of $F_4$ and $G_2$ is given by particular Fuchsian groups.   
 We show that the perturbation expansion of these integrated correlators is universal in the sense that it can be written as a single function of three parameters, called Vogel parameters, and a suitable 't Hooft-like coupling. To obtain the perturbative expansion for the integrated correlator with a given gauge group we simply need substituting in this single universal expression specific values for the Vogel parameters.
At the non-perturbative level we conjecture a formula for the one-instanton Nekrasov partition function valid for all simple gauge groups and for general $\Omega$-deformation background. We check that our expression reduces in various limits to known results and that it produces, via supersymmetric localisation, the same one-instanton contribution to the integrated correlator as the one derived from the lattice sum representation.
 Finally, we consider the action of the hyperbolic Laplace operator with respect to $\tau$ on the integrated correlators with exceptional gauge groups and derive inhomogeneous Laplace equations  very similar to the ones previously obtained for classical gauge groups.

\end{abstract}                                            
\end{center}

\newpage

\tableofcontents

\section{Introduction and outline}

In \cite{Binder:2019jwn} certain integrated correlators of four superconformal primary operators of the stress tensor multiplet in $\cN=4$ supersymmetric Yang--Mills (SYM) theory with gauge group $G$ have been derived starting from the work of Pestun \cite{Pestun:2007rz} on the $\cN=2^*$ SYM partition function written in terms of the Nekrasov partition function \cite{Nekrasov:2002qd}. 
Given that  $\cN=2^*$ SYM theory reduces to $\cN=4$ in the limit where $m\to 0$, with $m$ denoting the hypermultiplet mass, we can determine various $\cN=4$ quantities via supersymmetric localisation starting from Pestun partition function,  $Z_{G}(\tau,m)$, for $\cN=2^*$ SYM on a four-sphere.

In particular we here consider the quantity 
\bea
\cC_{G}(\tau):=\frac{1}{4} \Delta_\tau\partial_m^2 \log Z_{G}(\tau,m) \Big\vert_{m=0}\,,
\label{eq:firstcorr}
\eea
where $\Delta_\tau := 4 \tau_2^2\partial_\tau \partial_{\bar \tau}$ denotes the laplacian on the hyperbolic plane parametrised by the complexified $\cN=4$ coupling, $\tau= \tau_1+i\tau_2 := \frac{\theta}{2\pi} + i \frac{4\pi}{g_{_{Y\!M}}^2}$, with $\theta$ the theta angle and $g_{_{Y\!M}}$ the Yang--Mills coupling constant. 
In \cite{Binder:2019jwn} it was shown that this expression \eqref{eq:firstcorr} is proportional to a four-point correlator whose insertion points, $x_i$, are integrated over a measure factor $ \mu(x_1,\dots,x_4)$, schematically of the form 
 \bea
 \int \prod_{i=1}^4 dx_i \, \mu(x_1,\dots,x_4)\,\langle \cO_2(x_1) \dots \cO_2(x_4)\rangle\, ,
 \label{intcorr}
 \eea
 where $\cO_2(x)$ denotes the superconformal primary operator in the stress tensor supermultiplet, which is in the ${\bf 20^\prime}$ of the $SU(4)$ R-symmetry group. We refer to~\cite{Binder:2019jwn}  for a precise formulation of~\eqref{intcorr}. 
 
While for classical gauge groups, $G=SU(N),SO(2N),SO(2N+1),USp(2N)$, it is a relatively easy task to extract from the localised partition function, $Z_{G}(\tau,m)$, the perturbative expansion at weak coupling, $ g_{_{Y\!M}}^2\to0$, of the correlator for any value
of $N$ \cite{Chester:2019pvm, Chester:2020dja, Alday:2021vfb}, extracting the non-perturbative instanton contributions, which are contained in Nekrasov partition function, is far more involved.  

While for general gauge group we do not expect any simplification, in the case of classical gauge groups,~i.e.~$G=SU(N),SO(2N),SO(2N+1),USp(2N)$, we have the extra parameter, $N$, coming to the rescue. In these cases, the  problem notably simplifies at large-$N$ and, as shown in \cite{Chester:2019jas,Chester:2020vyz,Alday:2021vfb}, $SL(2,\Z)$ Montonen--Olive duality \cite{Montonen:1977sn}  (also known as S-duality) constraints the coefficients in the perturbative $1/N$ expansion at fixed $\tau$ to be sums of non-holomorphic, modular invariant Eisenstein series with half-integer index. 
 
 Surprisingly, in a series of papers \cite{Dorigoni:2021bvj,Dorigoni:2021guq, Dorigoni:2022zcr} an exact and modular covariant expression was conjectured and then proven for the integrated correlator \eqref{eq:firstcorr} with any classical gauge group and finite $\tau$.\footnote{See \cite{Dorigoni:2022iem} for a recent review, as well as \cite{Green:2020eyj, Dorigoni:2021rdo} for the extension to higher-point maximal $U(1)_Y$-violating correlators and \cite{Paul:2022piq,Brown:2023cpz,Paul:2023rka,Brown:2023why} for the generalisation to integrated four-point correlators involving operators with higher conformal dimensions and \cite{Pufu:2023vwo} for integrated two-point functions of superconformal primaries of the stress-energy tensor multiplet in the presence of a half-BPS line defect.}  
 In the present work, we revisit such results and show that for all simple gauge groups, classical and exceptional, the integrated correlator \eqref{eq:firstcorr} can be expressed for all values of $\tau$ via the exceptionally simple formula
  \bea
\tcboxmath{\cC_{G} (\tau)  = 
 \sum_{(m,n)\in\mathbb{Z}^2}   \int_0^\infty  \left[ e^{ - t\,  Y_{mn}(\tau)} B_{\mathfrak{g}}(t)  + e^{ - t\,  Y_{mn}(\mathfrak{n}_{\mathfrak{g}} \tau)} B_{\!\,^L\mathfrak{g}}(t)\right]  dt   \, ,
\label{gsun}}
\eea
where we have defined the quantity
\begin{equation}
Y_{mn}(\tau) :=\pi \frac{|m+n\tau|^2}{ \tau_2}\,.\label{eq:Ymn}
\end{equation} 

Since the correlator \eqref{eq:firstcorr} relates to a four-point function of local operators, it is really only sensitive to the simple Lie algebra, $\mathfrak{g}$, corresponding to the simple gauge group $G$, and its Goddard--Nuyts--Olive or Langlands dual $^L\mathfrak{g}$,~i.e.~we are not sensitive to global aspects of $G$ and $^LG$. In particular the quantity $\mathfrak{n}_\mathfrak{g}$ appearing in \eqref{gsun} denotes the ratio between the length square of the long and short roots of $\mathfrak{g}$.
The ``Borel transform'', $B_\mathfrak{g}(t)$, is a rational function of the form
 \begin{equation}
B_{\mathfrak{g}} (t)= \frac{\mathcal{Q}_{\mathfrak{g}}(t) }{ (t+1)^{2 h^\lor+1} }\,,\label{eq:BN}
 \end{equation}
 where $h^\lor$ is the dual Coxeter number of the Lie algebra $\mathfrak{g}$,  while $\mathcal{Q}_{\mathfrak{g}}(t)$ is a degree $2 h^\lor-1$ polynomial without constant term and with the ``palindromic'' property $\mathcal{Q}_{\mathfrak{g}}(t)=t^{2 h^\lor } \mathcal{Q}_{\mathfrak{g}}(t^{-1})$.
 
  For simply laced groups $G= SU(N)$, $SO(2N)$, $E_{6,7,8}$ the correlators are expected to be invariant under the $SL(2,\Z)$ action 
  \begin{equation}
  \tau \to \gamma\cdot \tau := \frac{a\tau+b}{c\tau+d}\qquad {\rm with} \qquad \gamma = \left(\begin{matrix}
a& b \\ c & d \end{matrix}\right)\in { SL}(2,\Z)\,,
\end{equation} which is a consequence of Montonen--Olive duality.
  In these cases, we have $\mathfrak{n}_\mathfrak{g}=1$ and $\mathfrak{g} =\! \!\,^L\mathfrak{g}$ so that 
  \begin{equation} 
  B_{\mathfrak{g}}(t)  = B_{^L\mathfrak{g}}(t)\qquad\qquad \rm{for}\,\, \mathfrak{g}= \{\mathfrak{su}_N,\,\mathfrak{so}_{2N},\, \mathfrak{e}_6,\,\mathfrak{e}_7,\, \mathfrak{e}_8\}\,,
  \end{equation}
  and equation \eqref{gsun} is manifestly invariant under $SL(2,\Z)$. 
  
 For non-simply laced groups we have to distinguish between the classical cases $G =SO(2N+1) $, $USp(2N)$  and the exceptional cases $G= F_4,G_2$. 
 Starting with $G =SO(2N+1)$, $USp(2N)$, we have that $\mathfrak{n}_\mathfrak{g}=2$ so that the integrated correlator \eqref{gsun} is only invariant under the congruence subgroup $\Gamma_0(2)\subset SL(2,\mathbb{Z})$.\footnote{The elements of the congruence subgroup $\Gamma_0(r)$ are given by $\gamma = \left(\begin{smallmatrix}
a& b \\ c & d \end{smallmatrix}\right)\in SL(2,\Z)$ with $c\equiv 0\,(\mbox{mod}\,r)$.} Furthermore, since the Lie algebras ${\mathfrak{so}_{2N+1}}$ and $\mathfrak{sp}_{2N}$ are Langlands duals to each others we have \cite{Dorigoni:2022zcr}
 \ie \label{eq:GNO}
 B_{\mathfrak{so}_{2N+1}}(t) = B_{^L\mathfrak{sp}_{2N}}(t) \,,\qquad\qquad B_{\mathfrak{sp}_{2N}}(t) = B_{^L\mathfrak{so}_{2N+1}}(t) \,,
 \fe
 which make Goddard--Nuyts--Olive (GNO) duality \cite{Goddard:1976qe} of \eqref{gsun}  manifest.  
 
The remaining exceptional groups $G=F_4$ and $G_2$ are Langlands self-dual hence we have
 \begin{equation}
  B_{\mathfrak{f}_4}(t) =   B_{^L\mathfrak{f}_4}(t)\,,\qquad\qquad\qquad B_{\mathfrak{g}_2}(t) =   B_{^L \mathfrak{g}_2}(t)\,.
 \end{equation}
 Furthermore, for these exceptional cases the integrated correlators \eqref{gsun} result invariant under the \textit{Hecke triangle group} (particular instance of   \textit{Fricke group}) $\Gamma_{(2,2\mathfrak{n}_\mathfrak{g},\infty)}$, where $\mathfrak{n}_\mathfrak{g}=2$ for $G=F_4$ and $\mathfrak{n}_\mathfrak{g}=3$ for $G=G_2$.  For $\mathfrak{n}_\mathfrak{g}=2,3$ the group $\Gamma_{(2,2\mathfrak{n}_\mathfrak{g},\infty)}$ is an infinite discrete subgroup of $SL(2,\mathbb{R})$ not isomorphic to $SL(2,\mathbb{Z})$ which extends the manifest $\Gamma_0(\mathfrak{n}_\mathfrak{g})$ symmetry of \eqref{gsun} and reproduces the expected Goddard--Nuyts--Olive (GNO) duality groups \cite{Girardello:1995gf,Dorey:1996hx,Argyres:2006qr,Kapustin:2006pk} for the non-simply laced exceptional gauge-groups $G=F_4$ and $G_2$.
 
A first consequence of our analysis concerns the small-coupling expansion, $g_{_{Y\!M}}^2\to0$ of the integrated correlator \eqref{eq:firstcorr}. We show that such perturbative integrated correlators in $\mathcal{N}=4$ SYM are \textit{universal} in the sense of Vogel  \textit{universal Lie algebra} \cite{VOGEL20111292,VogelPre}. From a mathematical point of view a universal Lie algebra can be seen as a certain tensor category with a moduli space, usually named \textit{Vogel plane}, given by the quotient space $\mathbb{P}^2/\mathfrak{S}_3$ of a three-dimensional projective plane with projective coordinates $(\alpha,\beta,\gamma)$, called Vogel parameters, quotiented by the action of the symmetric group $\mathfrak{S}_3$ on $(\alpha,\beta,\gamma)$. All simple Lie algebras arise as special points on the Vogel plane and we define universal quantities as  analytic, in the present case rational, functions on the Vogel plane. With a suitable (and universal) definition of the appropriate Yang-Mills coupling we show that the weak coupling expansion of \eqref{eq:firstcorr} is a universal quantity.
 
We then move to non-perturbative consequences of our result \eqref{gsun}.
While it is rather straightforward to extract the Yang-Mills instanton contributions from the conjectural expression \eqref{gsun}, matching such terms against supersymmetric localisation turns out to be rather hard.
From the supersymmetric localisation point of view, instantons corrections are fully captured by Nekrasov partition function which for classical gauge groups can be derived from an ADHM construction of the instanton moduli space \cite{Nekrasov:2002qd} or alternatively by exploiting the modular anomaly equation satisfied by the prepotential of $\mathcal{N}=2^*$ SYM \cite{Billo:2015pjb,Billo:2015jyt,Billo:2016zbf}. 

Neither of these constructions is available for the exceptional groups $G=F_4,G_2$. Heavily relying on the results of \cite{Billo:2015pjb,Billo:2015jyt,Billo:2016zbf}, we conjecture a very compact formula for the one-instanton Nekrasov partition function which depends solely on the roots of the associated Lie algebra. Our conjectural expression reproduces previous results derived in \cite{Billo:2015jyt} for $\mathcal{N}=2^*$ SYM with a non-simply laced gauge group in absence of $\Omega$-deformation. Furthermore in the hypermultiplet decoupling limit, $m\to\infty$, where $\mathcal{N}=2^*$ SYM reduces to pure $\mathcal{N}=2$ SYM theory, we find that our conjectural expression reproduces correctly the one-instanton partition function obtained in \cite{Keller:2011ek} for pure $\mathcal{N}=2$ SYM  on a non-trivial $\Omega$-deformation background. Finally, with this candidate exceptional Nekrasov partition function we compute from supersymmetric localisation the one-instanton contribution to the integrated correlator \eqref{eq:firstcorr} which matches beautifully with the conjectural exact lattice sum expression \eqref{gsun}, thus strengthening our claims.

Lastly, it was proven in~\cite{Dorigoni:2022zcr} that for all classical gauge groups the integrated correlators satisfy Laplace-difference equations that generalise the equation satisfied in the $SU(N)$ case originally discovered in~\cite{Dorigoni:2021bvj,Dorigoni:2021guq}.
A striking consequence of such equations is that, given the initial condition~$\cC_{SU(1)}(\tau)= 0$, all correlators~$\cC_{G}(\tau)$ for any classical gauge group,~$G$, are uniquely determined in terms of~$\cC_{SU(2)}(\tau)$. Although we did not manage to find a universal Laplace difference formula, we show that similar inhomogeneous Laplace equations are obeyed in the case of exceptional gauge groups as well. For all simple groups $G$ the integrated correlator $\cC_{G}(\tau)$ is uniquely determined in terms of~$\cC_{SU(2)}(\tau)$.
 
\subsection{Outline}

We start section~\ref{intgencorr} with a brief review of how the integrated correlator, $\C_{G}$, defined in \eqref{eq:firstcorr} can be computed by taking derivatives of the supersymmetric localised partition function of $\cN=2^*$ SYM on $S^4$ in the limit $m\to 0$. 
Firstly, we compute the perturbation expansion, $g_{_{Y\!M}}^2\to0$, of the integrated correlator where Yang-Mills instantons are suppressed. We show that such a perturbative series can be written for a universal $\cN=4$ gauge theory rather than for each specific and concrete example of gauge group $G$. Order by order in a suitable universal `t Hooft-like coupling, we find simple coefficients given by rational functions on the Vogel plane, which reduce to known expressions when we specialise to points on the Vogel plane corresponding to all classical Lie algebras.

We then review how to derive, using Nekrasov partition function \cite{Nekrasov:2002qd}, the contribution to \eqref{eq:firstcorr} coming from $G=SU(N)$ Yang-Mills instantons and how this can be generalised to arbitrary classical gauge groups \cite{Billo:2015pjb, Billo:2015jyt}. Based on the results of \cite{Billo:2015pjb,Billo:2015jyt,Billo:2016zbf}, we propose a simple expression for the one-instanton Nekrasov partition function on a non-trivial $\Omega$-deformation background which is valid for an arbitrary gauge group and depends only on the roots of the associated Lie algebra. Our conjectural formula is shown to reproduce known results when we specialise it to the flat-space case,~i.e. in absence of $\Omega$-deformation, and in the hypermultiplet decoupling limit, $m\to\infty$, where $\cN=2^*$ reduces to pure $\cN=2$ SYM. 

In section \ref{sec:ansatz}, we conjecture the exact non-perturbative expression for the integrated correlator, $\C_{G}(\tau)$, with arbitrary simple gauge group, $G$, given as the two-dimensional lattice sum in \eqref{gsun}.
We briefly review GNO duality for $\mathcal{N}=4$ SYM with arbitrary simple gauge group $G$. 
In particular, we show that, while \eqref{gsun} is manifestly $SL(2,\mathbb{Z})$ invariant under GNO duality for $G=E_{6,7,8}$, in the exceptional cases $G=F_4$, $G_2$ it is invariant under the respective GNO duality Hecke groups. 
Furthermore, for the novel cases of exceptional gauge groups $G=E_{6}$, $F_4$ and $G_2$, we show that this formula reproduces the expected perturbative expansion previously discussed and matches at the one-instanton level with the supersymmetric localisation results obtained from our conjectural exceptional Nekrasov partition function.   

We then consider in section \ref{sec:lapdiff} the action of the hyperbolic laplacian on the exact integrated correlator \eqref{gsun}. We show that in the cases where $G$ is exceptional the integrated correlator satisfies inhomogeneous Laplace equations similar to the ones discovered in \cite{Dorigoni:2021bvj,Dorigoni:2021guq,Dorigoni:2022zcr} for all classical gauge groups. Even for exceptional groups, $G$, the integrated correlator $\cC_{G}(\tau)$ is uniquely determined in terms of $\cC_{SU(2)}(\tau)$.
We conclude in section~\ref{sec:discussion} with a summary of these results and of possible future directions.

\section{Integrated correlators for general simple groups} 
\label{intgencorr}

The starting point of our analysis is the  $S^4$  partition function,~$ Z_{G}(m,\tau)$, of $\cN=2^*$ SYM with gauge group $G$ determined in~\cite{Pestun:2007rz} by Pestun using supersymmetric localisation and given by
\bea
 Z_{G}(m,\tau)  &:=&  \frac{1}  {\cN_{G}} \int v_{G}({a}) \, e^{-\frac{8\pi^2}{g_{_{Y\!M}}^2} \langle a,a\rangle_G} \, \hat Z_{G}^{pert}(m; a) \,  |\hat Z_{G}^{inst}  (m ,\tau; a)|^2 \,d^r \!a\nn\\
   &=& \langle \,   \hat{Z}_{G}^{pert}(m; a) \,  |\hat Z_{G}^{inst}  (m ,\tau ; a)|^2  \,  \rangle_{G}\,,
 \label{partfun}
 \eea
 where the integration variable, $a$, runs over the $r$-dimensional Cartan subalgebra of $G$, $v_{G}(a)$ is the Vandermonde determinant associated with the group $G$:
 \begin{equation}
 v_{G}(a) := \prod_{ \alpha \in \Delta} | (\alpha \cdot a) |= \prod_{\alpha \in \Delta^+} (\alpha \cdot a)^2 \,,\label{eq:vdm}
 \end{equation}  
 where $\Delta$ denotes the complete set of roots for $G$, while $\Delta^+$ denotes only the positive roots of $G$.
 In our normalisation the Killing form $\langle a,a\rangle_{G}$  is given by
 \begin{equation}
 \langle a,a\rangle_{G} := \mathfrak{n}_\mathfrak{g} \sum_{i=1}^r a_i^2\,,
 \end{equation}
 where for future reference it is convenient to define  the ratio
\begin{equation}
\mathfrak{n}_\mathfrak{g} := \frac{|\alpha_{\rm long}|^2 }{ |\alpha_{\rm short}|^2}\,,
\end{equation}
with $|\alpha_{\rm long}|^2$ and $|\alpha_{\rm short}|^2$ denoting respectively the length square of the long and short roots of $\mathfrak{g}$.
In particular we have
\begin{align}
\mathfrak{n}_\mathfrak{g} &\notag = 1\,, \qquad \mathfrak{g} = \{\mathfrak{su}_N,\,\mathfrak{so}_{2N},\, \mathfrak{e}_6,\,\mathfrak{e}_7,\, \mathfrak{e}_8\}\,,\\
\mathfrak{n}_\mathfrak{g}& =2\,,\qquad \mathfrak{g}= \{\mathfrak{so}_{2N+1},\,\mathfrak{sp}_{2N},\,\mathfrak{f}_4\}\,,\\
\mathfrak{n}_\mathfrak{g}&\notag = 3 \,,\qquad\mathfrak{g} = \mathfrak{g}_2\,,
\end{align}
which can be easily checked from the standard root systems summary presented in appendix \ref{app:Lie}.

 The normalisation factor $\cN_{G}$ is given by
 \ie
\cN_{G} :=  \int  v_{G}(a) \, e^{-\frac{8\pi^2}{g_{_{Y\!M}}^2} \langle a,a\rangle_G } \,d^r \!a\, .
 \fe
The matrix model expectation value of a general function $F(a)$ is defined by
 \bea
 \langle F(a)\rangle_{G} :=  \frac{1}{\cN_{G }} \int  v_{G}(a) \, e^{-\frac{8\pi^2}{g_{_{Y\!M}}^2} \langle a,a\rangle_G } \, F(a)\,d^r \!a\,,
 \label{measure}
 \eea
 so that, with the given definition for $\cN_{G}$ above, we have $\langle \,   1  \,  \rangle_{G}=1$. 
 
 The perturbative contribution to the partition function is one-loop exact and is given by the classical action factor 
 \begin{equation}
 \exp\Big(-\frac{8 \pi^2}{ g_{_{Y\!M}}^2}\langle a,a\rangle_G \Big)\,,\label{eq:Sclass}
 \end{equation}
  multiplied by the one-loop determinant
 \begin{align}
 \hat Z_{G}^{pert}(m;a) &\notag:=  \frac{1}{H(m)^r} \prod_{\alpha\in \Delta}\! \frac{ H(\alpha\cdot a) }  {\big[ H(\alpha \cdot a+ m)H(\alpha \cdot a- m)  \big]^{\frac{1}{2}}  }\\
 &\phantom{:}=\frac{1}{H(m)^r}  \prod_{\alpha \in \Delta^+} \frac{ H(\alpha\cdot a)^2 }{ H(\alpha \cdot a+ m)H(\alpha \cdot a- m) }\,.
\label{zpert}
\end{align}
The even function $H(z)$ is given by $H(z):=e^{-(1+\gamma)z^2}\, G(1+iz)\, G(1-iz)$, where  $G(z)$ is Barnes $G$-function and $\gamma$ is Euler constant.  

The factor  $|\hat Z_{G}^{inst}|^2 = \hat Z_{G}^{inst} \,\overline{{\hat Z}_{G}^{inst}}$ in \eqref{partfun} is the contribution from Nekrasov partition function which encapsulates the non-perturbative effects coming from instantons and anti-instantons localised at the north and south  poles of $S^4$. While this factor is not known in full generality for arbitrary group, $G$, we will provide in the next section a conjectural expression at the one-instanton level valid for all simple gauge groups.
    
We now use these results to compute the quantity of interest,
\bea
\cC_{G}(\tau):=\frac{1}{4} \Delta_\tau\partial_m^2 \log Z_{G}(\tau,m) \Big\vert_{m=0}\,,
\label{eq:firstcorrIT}
\eea
which, as already mentioned, was shown in \cite{Binder:2019jwn} to reproduce the correlator of four superconformal primary operators  of the stress tensor supermultiplet integrated over their positions with a specific measure that maintains supersymmetry.

To derive a useful expression for $\cC_{G}(\tau)$ from the localised partition function \eqref{partfun}, we first perform a Fourier mode decomposition with respect to $\tau_1 = \theta/(2\pi)$ and write
\bea
\C_{G} (\tau) = \C_{G}^{(0)} (\tau_2) + \sum_{k=1}^\infty  \left(e^{2\pi i k \tau}\, \C_{G}^{(k)}(\tau_2) +e^{-2\pi i k \bar\tau}\, \C_{G}^{(-k)}(\tau_2) \right) \, .
\label{fmodes}
\eea
We can then easily separate our analysis into perturbative and instanton contributions since
\ie
\partial^2_{m} \log Z_{G}  \big{|}_{m=0} = \partial^2_{m} \log Z_{G}^{pert}  \big{|}_{m=0} + \partial^2_{m} \log Z_{G}^{inst}  \big{|}_{m=0} \, ,
\fe
where each contribution can be expressed as an expectation value in a gaussian matrix model,   
\begin{align}
 \partial^2_{m} \log Z_{G}^{pert}  \big{|}_{m=0} &:= \langle \partial_m^2 \,\hat Z_{G}^{pert} \big{|}_{m=0} \rangle_{G}\, , \\
  \partial^2_{m} \log Z_{G}^{inst}  \big{|}_{m=0}  &:= \langle \partial_m^2 \vert \hat Z_{G}^{inst}\vert^2 \big{|}_{m=0} \rangle_{G} =  \langle \partial_m^2 \Big( \hat Z_{G}^{inst} +  \overline{{\hat{Z}}_{G}^{inst}}\Big) \big{|}_{m=0} \rangle_{G} \,,
 \end{align}
 using that ${\hat{Z}}_{G}^{inst}(m=0,\tau;a)=1$ and the fact that both $\hat{Z}_{G}^{pert}$ and $\hat{Z}_{G}^{inst} $ are even functions of $m$.
 
From the Fourier mode decomposition \eqref{fmodes} we then see that the $k=0$ term is the purely perturbative contribution,
\bea
\C_{G}^{pert} (\tau_2) :=\C_{G}^{(0)} (\tau_2) = \frac{1}{4} \tau_2^2 \partial_{\tau_2}^2 \,\partial^2_{m} \log Z_{G}^{pert}  \Big{|}_{m=0}  \,,
\label{propseries}
\eea
while the $k\ne 0$ terms are the instanton and anti-instanton contributions,
\bea
\C_{G}^{inst} (\tau)\  := \frac{1}{4}\Delta_\tau\,\partial^2_{m} \log Z_{G}^{inst}  \Big{|}_{m=0} = \sum_{k=1 }^\infty \Big( e^{2\pi i k \tau}\, \C_{G}^{(k)}(\tau_2)+e^{-2\pi i k \bar{\tau}}\, \C_{G}^{(-k)}(\tau_2)\Big) \,.
\label{cinst}
\eea
  Since the integrated correlator is real it follows that the $k$-instanton and $k$-anti-instanton contributions are identical,~i.e. $\C_{G}^{(k)}(\tau_2) = \C_{G}^{(-k)}(\tau_2)$. We will come back to the instanton sectors in section \ref{sec:yminst} where we discuss how these can be computed in supersymmetric localisation using Nekrasov partition function and presently focus on rewriting the Fourier zero-mode sector.
 
Given the expression for the one loop determinant \eqref{zpert}, we see that the perturbative part of the integrated correlator is entirely captured by
\begin{equation}
 \partial^2_{m} \log Z_{G}^{pert}  \big{|}_{m=0}  = \langle \partial_m^2 \hat Z_{G}^{pert} \big{|}_{m=0} \rangle_{G} = \langle 2 \sum_{\alpha \in \Delta^+} K'( \alpha \cdot a)\rangle_G \,,
 \end{equation} 
 where we defined
\begin{equation}
 K(z) := -\frac{H'(z)}{H(z)}\,.
\end{equation}
From the known \cite{Russo:2013kea} integral representation for $K'(z)$,
\begin{equation}
 K'(z)  = -\int_0^\infty \frac{2w [\cos(2wz)-1]}{\sinh^2{w}}dw = \sum_{k=1}^\infty 2(2k+1)\zeta_{2k+1} (-1)^{k+1} z^{2k}\,,
 \end{equation}
 we arrive at 
 \begin{equation} \label{eq:ZGpert}
 \partial^2_{m} \log Z_{G}^{pert}  \big{|}_{m=0}  =  \langle 2 \sum_{\alpha \in \Delta^+} K'( \alpha \cdot a)\rangle_G  =\sum_{k=1}^\infty 4(2k+1)\zeta_{2k+1} (-1)^{k+1}   \langle \sum_{\alpha \in \Delta^+}  ( \alpha \cdot a) ^{2k}\rangle_G\,,
 \end{equation} 
 where the gaussian matrix model expectation value, $\langle \dots \rangle_{G}$, is defined by \eqref{measure} and can be easily computed order by order in the above expansion for any simple Lie algebra using the data provided in appendix \ref{app:Lie}.

\subsection{Universal perturbative expansion}
\label{pertexpn}

\begin{table}
\centering
\begin{tabular}{ cc }   
 \begin{tabular}{|c| c| c| c|} 
 \hline
 Type & Lie algebra $\mathfrak{g}$ & dim$\,\mathfrak{g}$ & $h^\lor$ \\ [0.3ex] 
 \hline\hline
 $A_N$   & $\mathfrak{su}_{N+1}$  &  $N(N+2)$   & $N+1$  \\
$B_N$   & $\mathfrak{so}_{2N+1}$ &  $N(2N+1)$ & $2N-1$ \\
 $C_N$  & $\mathfrak{sp}_{2N}$   &   $N(2N+1)$   &   $ N+1$ \\
 $D_N$  & $\mathfrak{so}_{2N}$   &   $N(2N-1)$  & $2N-2$  \\
 $E_6$  & $\mathfrak{e}_{6}$     &    $78$          &    $12$        \\
 $E_7$  & $\mathfrak{e}_{7}$       &    $133$        &   $18$     \\
 $E_8$ & $\mathfrak{e}_{8 }$     &    $248$       & $30$  \\
  $F_4$  & $\mathfrak{f}_{4}$       &    $52$         &  $9$   \\
  $G_2$  & $\mathfrak{g}_{2}$      &   $14$          & $4$ \\[1ex] 
 \hline
 \end{tabular}
 &  \quad  \vspace{0.2cm}
 \begin{tabular}{|c| c| c| c| c|} 
 \hline
 Lie Algebra $\mathfrak{g}$ & $\alpha$ & $\beta$ & $\gamma$ \\ [0.3ex] 
 \hline\hline
 $\mathfrak{su}_{N}$   & $-2$  &  $2$   & $N$  \\
 $\mathfrak{so}_{2N+1}$   & $-2$ &  $4$ & $2N-3$ \\
 $\mathfrak{so}_{2N}$  & $-2$   &   $4$  & $2N-4$  \\
 $\mathfrak{so}_{n}$  &  $-2$   &   $4$   & $n-4$\\
  $\mathfrak{sp}_{2N}$  & $-2$   &   $1$   &   $ N+2$ \\
 $\mathfrak{e}_{6}$  & $-2$     &    $6$          &    $8$        \\
 $\mathfrak{e}_{7}$  & $-2$       &    $8$        &   $12$     \\
 $\mathfrak{e}_{8}$ & $-2$     &    $12$       & $20$  \\
  $\mathfrak{f}_{4}$  & $-2$       &    $5$         &  $6$         \\
  $\mathfrak{g}_{2}$  & $-2$      &   $\frac{10}{3}$  & $\frac{8}{3}$ \\ [1ex] 
 \hline
 \end{tabular}
 \vspace{0.2cm}
 
 \\ 
 (a)~Data for simple Lie algebras. & (b)~Vogel parameters for simple Lie algebras.

\end{tabular}
\caption{  } \label{tab:Lie}
\end{table}

 Before discussing the universal nature of $\mathcal{N}=4$ perturbative expansion, we briefly review the notion of Vogel universal Lie algebra. 
 
In \cite{VogelPre} Vogel, motivated by knot theory, defined a certain tensor category to try and construct a category of modules over a general Lie algebra, thus intended to be a model for a universal Lie algebra.
Given $\mathfrak{g}$ a complex simple Lie algebra, Vogel considered the symmetric square of the adjoint representation, $S^2\mathfrak{g}$, which decomposes into three irreducible representations. Fixing a particular choice of invariant bilinear form on $\mathfrak{g}$ and denoting by $2t$ the Casimir eigenvalue of the adjoint representation, we then have that each of these modules have Casimir eigenvalues $4t-2\alpha, 4t-2\beta, 4t-2\gamma$, which can be taken as the definition for Vogel parameters $(\alpha,\beta,\gamma)$.

Given that the choice of an invariant bilinear form on a simple Lie algebra is unique up to scalar multiples, we are naturally led to considering  $(\alpha,\beta,\gamma)$ as projective coordinates.
This leads to the notion of \textit{Vogel plane}, which is defined as the quotient space $\mathbb{P}^2/\mathfrak{S}_3$ with coordinates precisely given by $\alpha,\beta,\gamma$, i.e.
\begin{equation}
(\alpha,\beta,\gamma) \sim (k\alpha, k \beta, k\gamma) \sim \rho(\alpha,\beta,\gamma)\,,
\end{equation}
with $k\neq 0$ and for all permutations $\rho\in \mathfrak{S}_3$. 

Vogel showed that many Lie algebraic quantities, like the dimensions of such modules, are \textit{universal} in that they can be expressed as analytic functions on the Vogel plane, i.e. as projective and permutations invariant functions of $\alpha,\beta,\gamma$. 
Just to present two easy examples, for the quadratic Casimir eigenvalue we have
\begin{align} 
\label{dualcoxeter}
t  = \alpha+\beta+\gamma\,,
\end{align}
and for the dimension of $\mathfrak{g}$ (or equivalently the dimension of the adjoint representation) we have
\begin{equation}
{\rm dim}\, \mathfrak{g} = \frac{(\alpha-2t)(\beta-2t)(\gamma-2t)}{\alpha\beta\gamma}\,.\label{eq:dimg}
\end{equation}

Once we have a universal expression for a certain quantity, its particular value for a given simple Lie algebra can be obtained by working at a specific point on the \textit{Vogel plane}. Choosing what is called ``minimal bilinear form'' normalisation where $\alpha=-2$, we present in Table \ref{tab:Lie}-(b) the special points in Vogel plane corresponding to all simple Lie algebras. 
If we substitute the Vogel plane points presented in Table \ref{tab:Lie}-(b) into the universal expressions \eqref{dualcoxeter}-\eqref{eq:dimg}, it is rather easy to see that in this normalisation we have $t = \alpha+\beta+\gamma= h^\lor$, with $h^\lor$ the dual Coxeter number of $\mathfrak{g}$ and that the dimension of $\mathfrak{g}$ is correctly reproduced for all Lie algebras, as presented in Table \ref{tab:Lie}-(a). 

Equations \eqref{dualcoxeter}-\eqref{eq:dimg} are the first examples of universal formulae in the study of simple Lie algebras, while several more universal quantities have been found in the literature, see for instance \cite{landsberg2006universal,westbury2003invariant,mkrtchyan2012casimir,Avetisyan:2018yac,Avetisyan:2019jfk,Avetisyan:2020qhn}.
Importantly, the notion of universal Lie algebra has been applied to gauge-invariant quantities in gauge theories. 

In particular in \cite{Mkrtchyan:2012jh,Mkrtchyan:2013htk} it was shown that for Chern-Simons theory on a $3$-sphere the central charge, the partition function, and the expectation value of the unknotted Wilson loop in the adjoint representation can all be expressed as universal quantities, i.e. for a universal Chern-simons theory. This constitutes a fruitful playground for establishing and investigating dualities between theories, such as the Chern-Simons/topological strings duality \cite{Krefl:2015vna} and more general level/rank dualities.

We will now show that a similar phenomenon takes place when discussing the perturbative expansion for the $\mathcal{N}=4$ integrated correlators as defined in \eqref{propseries}. 
Similar to what happens in universal Chern-Simons theory, we start by considering the original localised on-shell action appearing in \eqref{eq:Sclass} and given by
$$
\frac{8\pi^2 }{g_{_{Y\!M}}^2} \langle a,a\rangle_G  \,.$$
In order to define a universal perturbative expansion for the integrated correlator we should not be fixing any particular choice for invariant bilinear form $\langle a,a\rangle_G$ on a simple Lie algebra $\mathfrak{g}$, thus impliying that the coupling constant $g_{_{Y\!M}}^2$ is not quite a good universal quantity, since under a rescaling of invariant bilinear form, i.e. under $t=\alpha+\beta+\gamma \to k t$, $g_{_{Y\!M}}^2$ has to change accordingly as to leave the action invariant.

We conclude that the universal $\mathcal{N}=4$ SYM theory must depend on the quotient moduli space $\mathbb{P}^3/\mathfrak{S}_3$ parametrised by the four coordinates 
\begin{equation}
(g_{_{Y\!M}}^{-2},\alpha,\beta,\gamma)\sim (kg_{_{Y\!M}}^{-2},k\alpha,k\beta,k \gamma)\sim (g_{_{Y\!M}}^{-2},\rho(\alpha,\beta,\gamma))\,,
\end{equation}
with $k\neq 0$ and where the symmetric group, $\rho\in \mathfrak{S}_3$, acts only on $(\alpha,\beta,\gamma)$. 
Let us work from now on in the minimal bilinear form normalisation convention presented in Table \ref{tab:Lie}-(b). 
To obtain a universal perturbation expansion we are then led to introduce a modified 't Hooft coupling
\begin{equation}
a_{\mathfrak{g}} :=\frac{ t}{4\pi^2} g_{_{Y\!M}}^2=  \frac{ h^\lor}{4\pi^2} g_{_{Y\!M}}^2\,,
\end{equation}
where $h^\lor$ is again the dual Coxeter number, presented in universal form in \eqref{dualcoxeter}.

Another important universal quantity to introduce is the central charge (or conformal anomaly) of $\mathcal{N}=4$ SYM with gauge algebra $\mathfrak{g}$, which is given by
\begin{equation}
c_{\mathfrak{g}} :=\frac{ {\rm dim}\,\mathfrak{g}}{4}\,,
\end{equation}
where ${\rm dim}\,\mathfrak{g}$ denotes the dimension of the Lie algebra, or equivalently the dimension of the adjoint representation given by the universal expression \eqref{eq:dimg}.

Both the modified 't Hooft coupling $a_{\mathfrak{g}} $ (unlike $g_{_{Y\!M}}^2$) and the central charge $c_{\mathfrak{g}} $ are indeed universal functions defined on the quotient moduli space $\mathbb{P}^3/\mathfrak{S}_3$, i.e. they are invariant under rescaling $(g_{_{Y\!M}}^{-2},\alpha,\beta,\gamma)\sim (kg_{_{Y\!M}}^{-2},k\alpha,k\beta,k \gamma)$, with $k\neq 0$, and under permutations of $(\alpha,\beta,\gamma)$.

If we compute the perturbation expansion \eqref{propseries} for the $\mathcal{N}=4$ SYM integrated correlators expressed as a power series in the modified 't Hooft coupling, $a_{\mathfrak{g}} $, we find the universal perturbative expansion 
\begin{align}
&\nn\C_{G}^{pert} (\tau_2) = \\
&\nn  4 c_{\mathfrak{g}} \left[ \frac{3   \, \zeta_3\,a_{\mathfrak{g}}   }{2} -\frac{75 \, \zeta_5 \,a_{\mathfrak{g}}^2}{8} 
+\frac{735 \,\zeta_7 \,a_{\mathfrak{g}}^3}{16} -\frac{6615  \,\zeta_9  \left(1 + P_4({\mathfrak{g})}\right)  a_{\mathfrak{g}}^4 }  {32} +\, \frac{114345 \,  \zeta_{11} \left(1+  P_5({\mathfrak{g})}  \right)a_{\mathfrak{g}}^5  }{128 }\right. \\
& \qquad \left. 
 -\frac{3864861 \,\zeta_{13} \left(1 +  P_6({\mathfrak{g})}  \right) a_{\mathfrak{g}}^6}{1024}+\frac{32207175\,\zeta_{15} \left(1 +  P_7({\mathfrak{g})}  \right) a_{\mathfrak{g}}^7}{2048}+ O(a_{\mathfrak{g}}^{8}) \right] \,,
\label{pertexp}
\end{align}
where the ``non-planar'' factors $P_i(\mathfrak{g}) = P_i(\alpha,\beta,\gamma) $ are universal functions on Vogel plane $\mathbb{P}^2/\mathfrak{S}_3$, i.e.
$$
P_i(\mathfrak{g})=P_i( \alpha,\beta,\gamma ) = P_i(k \alpha,k\beta,k\gamma )  = P_i(\rho( \alpha,\beta,\gamma))\,,
$$
for all $k \neq 0$ and for all permutations $\rho \in \mathfrak{S}_3$.

Remarkably, if we assume that $P_i(\alpha,\beta,\gamma)$ is a permutation invariant, rational function of the projective Vogel parameters with denominator given by $t^{i-1}$, or equivalently  ${(h^\lor)}^{i-1}$ (as suggested by the $SO(n)$ results derived in \cite{Dorigoni:2022zcr}) then the $SU(N)$ and $SO(2N)$ perturbation expansions uniquely fix $P_i(\alpha,\beta,\gamma)$ for all $i\leq 9$. If we furthermore input as well the $G_2$ and $F_4$ perturbative series, we can uniquely fix the coefficients $P_{i}(\alpha,\beta,\gamma)$ up to $i\leq 12$.

For the first few orders we explicitly find
\begin{align}
&\nn P_1(\mathfrak{g}) =P_2(\mathfrak{g}) =P_3(\mathfrak{g}) =0\,,\\
&\nn P_4(\mathfrak{g})  = -\frac{2 \sigma_1^3-3\sigma_1 \sigma_2 +\sigma_3}{84 \sigma_1^3} \,,\\
&\label{eq:uniP7}P_5(\mathfrak{g})  = \frac{7}{2}P_4(\mathfrak{g}) =-\frac{2 \sigma_1^3-3 \sigma_1\sigma_2 +\sigma_3}{24 \sigma_1^3} \,,\\
&\nn P_6(\mathfrak{g})   =-\frac{96 \sigma_1^5-145 \sigma_1^3\sigma_2 +51\sigma_1^2 \sigma_3 -3\sigma_1 \sigma_2^2+\sigma_2 \sigma_3}{528 \sigma_1^5}\,,\\
&\nn P_7 (\mathfrak{g}) =\frac{-10776 \sigma_1^6+16201 \sigma_1^4\sigma_2 -6171\sigma_1^3 \sigma_3 +1245 \sigma_1^2 \sigma_2^2-541 \sigma_1\sigma_2 \sigma_3+42 \sigma_3^2}{34320 \sigma_1^6}\,,
\end{align}
where 
\begin{equation} \sigma_1 := t= \alpha+\beta+\gamma, \qquad \qquad  \sigma_2 := \alpha^2 +\beta^2 +\gamma^2\,,\qquad \qquad \sigma_3 := \alpha^3 +\beta^3 +\gamma^3\,.
\end{equation} Note that each coefficient, $P_i(\mathfrak{g})$, is indeed defined on the projective and permutation symmetric Vogel parameters.
In appendix \ref{app:UniPert} we present all $P_{i}(\alpha,\beta,\gamma)$ up to $i\leq 12$ and furthermore rewrite all of these terms in the alternative basis of symmetric polynomials in $3$-variables, 
\begin{equation}
t=\alpha+\beta+\gamma\,,\qquad \qquad s:=\alpha\beta+\alpha\gamma+\beta\gamma\,,\qquad \qquad p:=\alpha\beta\gamma\,.
\end{equation}

 As argued in  \cite{Dorigoni:2022zcr}, the reason for referring to the factors $P_i(\mathfrak{g}) $ as ``non-planar''  is that  in the cases corresponding to classical gauge groups, $\mathfrak{g}\in \{\mathfrak{su}_N, \mathfrak{so}_{2N}, \mathfrak{so}_{2N+1},\mathfrak{sp}_{2N}\}$, these non-trivial factors, appearing from four loops onward\footnote{In \cite{Dorigoni:2022zcr} the authors denoted the first non-trivial non-planar correction coefficient by $P_1(\mathfrak{g})$ rather than the current convention $P_4(\mathfrak{g})$. We find it more convenient for later purposes, to start our counting from $O(a_{\mathfrak{g}})$ and denote by $P_1(\mathfrak{g})$ the corresponding (and vanishing) non-planar correction.}, are suppressed in $1/N$ at large-$N$.

It is straightforward to show that if we substitute the Vogel parameters reported in Table \ref{tab:Lie}-(b)  for the classical Lie algebras, $\mathfrak{g}\in \{\mathfrak{su}_N, \mathfrak{so}_{2N}, \mathfrak{so}_{2N+1},\mathfrak{sp}_{2N}\}$, in the universal non-planar corrections \eqref{eq:uniP7}, we reproduce exactly the results of \cite{Dorigoni:2022zcr}.
For concreteness we have:
\begin{itemize} 
\item $SU(N)$, Vogel parameters $(\alpha,\beta,\gamma)_{\mathfrak{su}_N} = (-2,2,N)$:
\begin{equation}\label{suparam}
\begin{aligned}
P_4(\mathfrak{su}_N)  & =  \frac{2}{7N^2} \, , \qquad\qquad\qquad\qquad P_5(\mathfrak{su}_N) =  {1 \over N^{2}} \, , \\
 P_6(\mathfrak{su}_N) &=  \frac{25N^2 +4}{11 N^4}\,,  \qquad\qquad \qquad \,\,P_7(\mathfrak{su}_N) = \frac{605 N^2+332}{143 N^4}\,.
\end{aligned}
\end{equation}

Remembering that the universal $\mathcal{N}=4$ SYM is defined by the coordinates $(g_{_{Y\!M}}^{-2},\alpha,\beta,\gamma)$ on the quotient moduli space $\mathbb{P}^3/\mathfrak{S}_3$, we see that for $\mathfrak{su}_n$ we have
$$(g_{_{Y\!M}}^{-2},\alpha,\beta,\gamma)\Big\vert_{\mathfrak{su}_N}= (g_{_{Y\!M}}^{-2},-2,2,N)\,,$$
 which is equivalent to 
 $$ (g_{_{Y\!M}}^{-2},\alpha,\beta,\gamma)\Big\vert_{\mathfrak{su}_N}  \sim (-g_{_{Y\!M}}^{-2}, - \beta,- \alpha,- \gamma)\Big\vert_{\mathfrak{su}_N}= (-g_{_{Y\!M}}^{-2},-2,2,-N) =  (-g_{_{Y\!M}}^{-2},\alpha,\beta,\gamma)\Big\vert_{\mathfrak{su}_{-N}}\,.$$
We have thus re-obtained the formal equivalence $\mathfrak{su}_{-N} = \mathfrak{su}_{N}$ for which the correlator is invariant if we furthermore implement the rescaling $ g_{_{Y\!M}}^2 \leftrightarrow - g_{_{Y\!M}}^2$ so that 
$$a_{\mathfrak{su}_{-N}}\big\vert_{-g_{_{Y\!M}}^2} = a_{\mathfrak{su}_{N}}\big\vert_{g_{_{Y\!M}}^2}\,.$$
\item $SO(n)$\footnote{We noticed a typo in equation (2.8) of \cite{Dorigoni:2022zcr} where the denominator of $ P_5(\mathfrak{so}_n)$ ($P_{SO(n),2}$ in the reference) is off by a factor of two.} with $n=2N$ or $n=2N+1$: Vogel parameters $(\alpha,\beta,\gamma)_{\mathfrak{so}_n} = (-2,4,n-4)$
\begin{align}
P_4(\mathfrak{so}_n)  &\nn =  -\frac{n^2-14 n+32}{14 (n-2)^3}\, , \qquad\qquad\qquad \qquad \,P_5(\mathfrak{so}_n) =  -\frac{n^2-14 n+32}{{  4} (n-2)^3}\, \,, \\
P_6(\mathfrak{so}_n)  &\label{soparam}= - \frac{12 n^4-221n^3+1158 n^2-2432 n +1856}  {22 (n-2)^5} \, ,\\
P_7(\mathfrak{so}_n)  &\nn = - \frac{2(342 n^5 -7217 n^4 -48841 n^3 -153938 n^2 +239232n-149920)}{715(n-2)^6}\,.
\end{align}

\item $USp(n)$ with $n=2N$, Vogel parameters $(\alpha,\beta,\gamma)_{\mathfrak{sp_n}} = (-2,1,\frac{n+4}{2})$:
\begin{align}
P_4(\mathfrak{sp_n})  &\nn=  \frac{n^2+14 n+32}{14 (n+2)^3}\ \, ,\quad\qquad\qquad \qquad\qquad P_5(\mathfrak{sp_n}) =  \frac{n^2+14 n+32}{8 (n+2)^3} \, , \\
P_6(\mathfrak{sp_n}) &\label{uspparam}=  \frac{12 n^4 + 221n^3+1158 n^2 +2432 n +1856}  {22 (n+2)^5}\, ,\\
P_7(\mathfrak{sp}_n)  & \nn=  \frac{2(342 n^5 +7217 n^4 + 48841 n^3 + 153938 n^2 +239232n+149920)}{715(n+2)^6}\,.
\end{align}
\end{itemize}
Once more, we can exploit the universality of $\mathcal{N}=4$ SYM by using $\mathbb{P}^3/\mathfrak{S}_3$ coordinates $(g_{_{Y\!M}}^{-2},\alpha,\beta,\gamma)$ and consider the special points on the extended Vogel plane
\begin{equation}
(g_{_{Y\!M}}^{-2},\alpha,\beta,\gamma)\Big\vert_{\mathfrak{so}_n}= (g_{_{Y\!M}}^{-2},-2,4 , n-4)\,, \qquad (g_{_{Y\!M}}^{-2},\alpha,\beta,\gamma)\Big\vert_{\mathfrak{sp}_n}= (g_{_{Y\!M}}^{-2},-2,1 ,\frac{n+4}{2})\,.
\end{equation}
We then have the equivalence,
$$
(g_{_{Y\!M}}^{-2},\alpha,\beta,\gamma)\Big\vert_{\mathfrak{so}_n}\!\!\sim\! (\shortminus\frac{g_{_{Y\!M}}^{-2}}{2} ,\shortminus\frac{\beta}{2},\shortminus\frac{\alpha}{2},\shortminus\frac{\gamma}{2} )\Big\vert_{\mathfrak{so}_n} \!\!=\! \Big ((\shortminus2g_{_{Y\!M}}^{2})^{-1},\shortminus2,1,\frac{\shortminus n+4}{2}\Big) \!=\! ((\shortminus2g_{_{Y\!M}}^{2})^{-1},\alpha,\beta,\gamma)\Big\vert_{\mathfrak{sp}_{-n}}\,,
$$
manifesting the formal equality $ \mathfrak{so}_{n}= \mathfrak{sp}_{-n} $ when we combine it with the coupling constant change $g_{Y\!M}^2 \leftrightarrow -2 g_{Y\!M}^2$ as to preserve
$$
a_{\mathfrak{so}_{n}}\big\vert_{g_{Y\!M}^2} = a_{\mathfrak{sp}_{-n}}\big\vert_{-2 g_{Y\!M}^2} \,. 
$$

Finally, we note that equation \eqref{pertexp}, can be seen as a modified generating series for the expectation values 
$$\langle \sum_{\alpha\in \Delta+} (\alpha\cdot a)^{2k}\rangle\,,$$
as one can see from equation \eqref{eq:ZGpert}.
To clarify the statement, we first notice that for $SU(N)$ we have $a_{\mathfrak{su}_N} = \lambda /(4\pi^2)$ with $\lambda:= N g_{_{Y\!M}}^2$ the standard 't Hooft coupling. For $SU(N)$ the leading order in the large-$N$ limit can be easily obtained \cite{Dorigoni:2021guq} by setting to zero all non-planar corrections, $P_k(\mathfrak{su}_N)=O(N^{-1})$, so that \eqref{pertexp} reduces to the leading genus-zero planar correction $ \C^{(0)}_{SU}(\lambda)$,
\begin{align}
&\C_{SU(N)}^{pert} (\tau_2) \sim N^2 \C^{(0)}_{SU}(\lambda) =N^2 \sum_{k=1}^\infty  \frac{(-4)^{k+1}\Gamma(k+\frac{3}{2})^2}{\pi \Gamma(k)\Gamma(k+3)} \zeta_{2k+1}\Big(\frac{ \lambda}{4\pi^2}\Big)^k\\
&\notag =N^2 \Big[ \frac{3 \, \zeta_{3} }{2} \Big(\frac{ \lambda}{4\pi^2}\Big) -\frac{75 \, \zeta_{5}}{8} \Big(\frac{ \lambda}{4\pi^2}\Big)^2
+\frac{735 \,\zeta_{7} }{16} \Big(\frac{ \lambda}{4\pi^2}\Big)^2  -\frac{6615  \,\zeta_{9}  }  {32}\Big(\frac{ \lambda}{4\pi^2}\Big)^4  +O(\lambda^5)\Big]\,.
\end{align}

We can then combine this planar limit result with the supersymmetric localisation formula \eqref{eq:ZGpert}, to write a somewhat explicit expression for the general non-planar corrections factors $P_k(\mathfrak{g})$,
\begin{equation}\label{eq:PtoA}
{\rm dim}(\mathfrak{g}) \, \big[ 1+ P_k(\mathfrak{g})\big] = \frac{\pi \,\Gamma(k+2)\Gamma(k+3)}{4^{k+1}\,\Gamma(k+\frac{1}{2})\Gamma(k+\frac{3}{2})} A_{2k}(\mathfrak{g})\,,\qquad k\in \mathbb{N}^{>0}\,,
\end{equation}
where we defined
\begin{equation}\label{eq:Amom}
A_{k}(\mathfrak{g}):= \langle \sum_{\alpha\in \Delta} (\alpha\cdot a)^k \rangle_{G}\Big\vert_{a_{\mathfrak{g}=1}} = \frac{1}{\mathcal{N}_{G}\vert_{a_{\mathfrak{g}=1}}} \int  v_G(a) \,e^{-2 h^\lor \langle a,a\rangle_G} \Big[\sum_{\alpha\in \Delta}  (\alpha\cdot a)^k \Big]\,d^r\! a\,.
\end{equation}
Note that $A_0 (\mathfrak{g})=|\Delta| $ is simply the number of roots for the algebra $\mathfrak{g}$, however this term does not contribute to our perturbative expansion \eqref{pertexp} since it is killed by the action of $\tau_2^2\partial_{\tau_2}^2$. Furthermore, from \eqref{eq:Amom} it is clear that $A_{2k+1}(\mathfrak{g})=0$ while we can check by direct calculation that 
\begin{equation}
A_{2}(\mathfrak{g}) = \frac{ {\rm dim}(\mathfrak{g})}{2}\,,\qquad A_{4}(\mathfrak{g}) = \frac{5\, {\rm dim}(\mathfrak{g})}{8}\,,\qquad A_{6}(\mathfrak{g}) =\frac{35\, {\rm dim}(\mathfrak{g})}{32}\,.
\end{equation}
By plugging these expressions in \eqref{eq:PtoA} we find that the first three ``non-planar'' corrections are indeed vanishing,~i.e. $P_k(\mathfrak{g})=0$ for $k=1,2,3$.

The non-planar corrections are thus all encoded in the generating series,
\begin{equation}
\mathcal{A}(\mathfrak{g};x) := \sum_{k=0}^\infty A_k(\mathfrak{g}) \frac{x^k}{k!} =  \frac{1}{\mathcal{N}_{G}\vert_{a_{\mathfrak{g}=1}}} \int  v_G(a) \,e^{-2 h^\lor \langle a,a\rangle_G} \sum_{\alpha\in \Delta}\big[\exp( x \,\alpha\cdot a\big)\big]\,d^r\!a\,. \label{eq:generating}
\end{equation}
Even with the plethora of high-order corrections reported in appendix \ref{app:UniPert}, we did not succeed in evaluating this generating series in closed form. Of course there are infinitely many generating series which one may construct from the coefficients \eqref{eq:Amom}, however, our particular choice seems ``nice enough'' to hopefully lead to an exact and universal formula for $\mathcal{A}(\mathfrak{g};x)$ and thus for the whole perturbative expansion  \eqref{pertexp}  via \eqref{eq:PtoA}.

\subsection{Nekrasov partition function}
\label{sec:yminst}

We now turn our attention to the Yang-Mills instanton sector.
As already discussed in \cite{Dorigoni:2022zcr}, the instanton contributions to $\C_{G}$ can be evaluated from 
\begin{align}
\C_{G}^{inst}(\tau)=\frac{1}{4}\Delta_\tau\,\partial_m^2 Z_{G}^{inst}(m,\tau)\Big\vert_{m\rightarrow 0} \,,
\end{align}
where $Z^{inst}_{G}(m,\tau)$ is the $\mathcal{N}=2^*$ SYM instanton partition function, which can be computed by a matrix model integral over the variables $a_i$ of the Nekrasov partition function $\hat{Z}^{inst}_{G}(m,\tau;a_i)$.

Given the Fourier mode decomposition \eqref{cinst}, we can furthermore specify directly the $k$-instanton contribution,
\begin{equation}\label{eq:kinst}
 e^{2\pi i k \tau}\, \C_{G}^{(k)}(\tau_2) =  \frac{1}{4}\Delta_\tau \langle \partial_m^2  \,\hat{Z}_{G}^{inst,\,k}  \Big{|}_{m=0} \rangle_{G}\,,
\end{equation}
where the $k$-instanton Nekrasov partition function takes the form:
\begin{equation}
\hat{Z}_{G}^{inst,\,k} (m, \tau; a) = e^{2\pi i k \tau} \hat{Z}_G^{k}(m; a)\,,
\end{equation}
and we have assumed without loss of generality that $k>0$, since the anti-instanton sector is simply given by $\C_{G}^{(-k)}(\tau_2) = \C_{G}^{(k)}(\tau_2)$.

For example, the case $G=SU(N)$ was derived in \cite{Pestun:2007rz,Nekrasov:2002qd,Nekrasov:2003rj} and the one-instanton Nekrasov partition function for $\mathcal{N}=2^*$ SYM on $S^4$ is given by 
\begin{equation}
\hat{Z}_{SU(N)}^{k=1}(m; a) = -m^2 \sum_{\ell=1}^N \prod_{j\neq \ell} \frac{(a_\ell-a_j+i)^2 -m^2}{(a_\ell -a_j) (a_\ell -a_j+2i)}\,.
\end{equation}

Although not manifest in our notation, we stress that Nekrasov partition function here considered, $\hat{Z}_{G}^{k}$, refers to $\mathcal{N}=2^*$ SYM theory defined on a four-sphere, hence $\hat{Z}_{G}^{k}$ really denotes Nekrasov partition function computed in a non-trivial $\Omega$-deformation background corresponding to $S^4$. In~\cite{Nekrasov:2002qd,Nekrasov:2003rj}, a formula for $\hat{Z}_{SU(N)}^{k}$ in a general $\Omega$-deformation background, parametrised by $\epsilon_1,\epsilon_2$, was derived starting from
the ADHM construction of the moduli space of $SU(N)$ instantons, and therefore generalisable to the other classical cases $G=SO(2N),$ $SO(2N+1)$ and $USp(2N)$ \cite{Billo:2015pjb,Billo:2015jyt}. The four-sphere partition function can then be recovered as the special case $\epsilon_1=\epsilon_2=1$.

However, for exceptional gauge groups an ADHM construction of the instanton moduli space is not available and for these cases the method of equivariant localization cannot be applied to evaluate the instanton partition function. An alternative procedure was derived in \cite{Billo:2015pjb,Billo:2015jyt,Billo:2016zbf}, by exploiting the modular anomaly equation satisfied by the prepotential of $\mathcal{N}=2^*$ SYM. In particular, by solving this anomaly equation order by order in a small-mass expansion, expressions for the instanton prepotential were found in terms of modular forms of $\tau$ and of functions of the root system of $\mathfrak{g}$, allowing for a unified treatment of all Lie algebras. Specifically, for one-instanton contributions these expressions drastically simplify leading to very compact formulae which depend solely on the roots of the gauge algebras.

 In \cite{Billo:2015pjb} an expression for one-instanton contributions with simply laced algebras (ADE) was derived in a general $\Omega$-deformation background, whereas in \cite{Billo:2015jyt} non-simply laced algebras were considered and a one-instanton formula was explicitly found only in the undeformed gauge theory,~i.e. only for flat space. We now review the results obtained in the case of ADE groups and then build on the analysis of \cite{Billo:2015pjb,Billo:2015jyt,Billo:2016zbf} to conjecture an expression for the one-instanton Nekrasov partition function in a non-trivial $\Omega$-deformation background and valid for general gauge group $G$, in particular filling the gap for the previously unknown cases of $G_2$ and $F_4$.
\vspace{0.3cm}

\textit{Notation:} Not to clutter our notation, for the rest of this section we drop the index $G$ from the $k$-instanton Nekrasov partition function, $\hat{Z}_G^{k}$, and replace it in favour of an explicit dependence on the particular $\Omega$-deformation background considered. In particular, $\hat{Z}^{k}_{\mathbb{R}^4}$ denotes the flat-space, i.e. with $\Omega$-deformation parameters $\epsilon_1=\epsilon_2=0$, $k$-instanton Nekrasov partition function in  $\cN=2^*$ SYM, while $\hat{Z}^{k}_{\Omega}$ denotes the $k$-instanton Nekrasov partition function for generic $\epsilon_1,\epsilon_2$ $\Omega$-deformation background. Lastly, $\hat{Z}^{k}_{S^4}$ denotes the sought-after $k$-instanton Nekrasov partition function on $S^4$,~i.e. for $\epsilon_1=\epsilon_2=1$.
\vspace{0.3cm}

Given a simple Lie group $G$, we denote by $\Delta$ the associated root system and by $\Delta_L$ the long roots, furthermore for any root $\alpha \in \Delta$ we define the set
\begin{align}
\Delta(\alpha):=\{\beta \in \Delta\,\,\vert\,\,(\alpha^\lor \cdot \,\beta)=1 \}\,,\label{eq:Deltaalpha}
\end{align}
with
\begin{equation}
\alpha^\lor :=\frac{2}{(\alpha\cdot\alpha)}\alpha\,.
\end{equation}
Following \cite{Billo:2015pjb,Billo:2015jyt,Billo:2016zbf}, we then define the auxiliary function
 \begin{align}
\label{Gfunction}
\Gnek(x,\epsilon\,; a):=\sum_{\alpha\in\Delta_L}\frac{1}{(\alpha\cdot a)(\alpha\cdot a+\epsilon)} \prod_{\beta\in\Delta(\alpha)}\biggl(1+\frac{x}{\beta\cdot a}\biggr)\,,
\end{align}
which is a polynomial in $x$ of degree $2h^\lor -4$.

Relying heavily and building on the analysis of \cite{Billo:2015pjb,Billo:2015jyt,Billo:2016zbf}, we conjecture that the one-instanton Nekrasov partition function for a general simple Lie group $G$ and for generic $\Omega$-deformation background, $\epsilon_1,\epsilon_2$, is given by 
\begin{align}
\label{Zinstade}
\hat{Z}^{k=1}_\Omega(m; a)= &-\frac{1}{2\epsilon_1\epsilon_2}\biggl(m^2-\frac{\epsilon^2}{4}\biggr)\biggl(m^2-\frac{\epsilon^2}{4}+\epsilon_1\epsilon_2\biggr)\\
&\notag \times \sum_{n=0}^{2h^\lor-4} \frac{\epsilon^n}{n!} \biggl[\mathcal{E}_n\biggl(\frac{1}{2}+\frac{m}{\epsilon} \biggr)+\mathcal{E}_n\biggl(\frac{1}{2}-\frac{m}{\epsilon} \biggr) \biggr] \biggl(\frac{\partial}{\partial x}\biggr)^n \Gnek(x,\epsilon\,; i a )\Big\vert_{x=0}\,,
\end{align}
with $\epsilon:=\epsilon_1+\epsilon_2$ and where $\mathcal{E}_n(x)$ are Euler polynomials which can be easily constructed from the generating series
\begin{equation}
\frac{2 e^{ x t}}{e^t+1} = \sum_{n=0}^\infty \mathcal{E}_n(x) \frac{t^n}{n!}\,. 
\end{equation}

To justify our claim we now proceed to specialise equation \eqref{Zinstade} to particular limiting cases as to reproduce known results in the literature.
Firstly we can easily compute the flat-space limit, $\epsilon_1,\epsilon_2\to0$, for which equation \eqref{Zinstade}  reduces to
\begin{align}
\hat{Z}^{k=1}_{\mathbb{R}^4}(m; a)&= \notag -\frac{m^4}{\epsilon_1\epsilon_2}  \sum_{n=0}^{2h^\lor -4} \frac{m^{2n}}{(2n)!} \biggl(\frac{\partial}{\partial x}\biggr)^{2n} \Gnek(x,\epsilon\,; ia)\Big\vert_{\begin{smallmatrix} \!\!\!\!\!\!\!\!\!\!\!\! x=0\\ \epsilon_1=\epsilon_2=0\end{smallmatrix}} =-\frac{m^4}{\epsilon_1\epsilon_2}  \Gnek(m,\epsilon=0\,;ia)\\
&\label{eq:ZnekflatNSL}=-\frac{m^4}{\epsilon_1\epsilon_2}\sum_{\alpha\in\Delta_L}\frac{1}{(\alpha\cdot ia)^2} \prod_{\beta\in\Delta(\alpha)}\biggl(1+\frac{m}{\beta\cdot ia}\biggr)\,.
\end{align}
We note that in the special case $\epsilon_1=\epsilon_2=0$ the function $\Gnek(x,\epsilon{=}0;a)$ becomes an even polynomial in $x$ given that from \eqref{Gfunction} it is manifest that we have contributions from both $\beta\in \Delta(\alpha)$ and $\shortminus\beta\in \Delta(\shortminus\alpha)$ since clearly $\alpha \in \Delta_L$ if and only if  $\shortminus \alpha \in \Delta_L$.

When $G$ is non-simply laced, this expression \eqref{eq:ZnekflatNSL} reproduces precisely the results derived\footnote{The factor of $i$ multiplying $a$ stems from a different convention in the parametrisation of the Cartan subalgebra. To properly compare with \cite{Billo:2015pjb,Billo:2015jyt,Billo:2016zbf} we need to perform the replacement $a_{\rm there}=i a_{\rm here}$.} in \cite{Billo:2015jyt} for the one-instanton partition function in the flat-space $\mathcal{N}=2^*$ undeformed theory.
While for $G$ simply laced, we firstly note that in this case $\Delta_L = \Delta$, i.e. all roots are ``long'', and secondly we can easily check that $(\alpha\cdot\alpha)=2$ using the Lie algebras conventions summarised in appendix \ref{app:Lie}. Hence for $G$ simply laced \eqref{eq:ZnekflatNSL} can be rewritten as
\begin{align}
\hat{Z}^{k=1}_{\mathbb{R}^4}(m;a)=-\frac{m^4}{\epsilon_1\epsilon_2}\sum_{\alpha\in\Delta}\frac{1}{(\alpha\cdot ia)^2} \prod_{ (\alpha \cdot\beta)=1 }\biggl(1+\frac{m}{\beta\cdot ia}\biggr)\,,\label{eq:ZnekflatSL}
\end{align}
which again reproduces identically the results of \cite{Billo:2015pjb} for the one-instanton partition function of flat-space $\mathcal{N}=2^*$ SYM with ADE gauge group.

Moreover as a second independent check, we can consider \eqref{Zinstade} in the limit $m\rightarrow \infty$ where the hypermultiplet of $\cN=2^*$ becomes infinitely massive thus decoupling from the theory which then reduces to pure $\cN=2$ SYM.
To properly recover pure $\cN=2$ SYM, we need to consider the double scaling limit $m\rightarrow \infty$ and simultaneously $ q = e^{2\pi i \tau}\to 0$ while the $\cN=2$ SYM strong coupling scale $\Lambda^2 = m^2 \exp( \frac{2\pi i \tau}{h^\lor}) $ is kept fixed\footnote{We note that the factor $h^\lor$ at exponent is precisely the one-loop $\beta$-function coefficient for pure $\cN=2$ SYM with gauge group $G$. The strong-coupling scale definition $\Lambda^2 = m^2 \exp( \frac{2\pi i \tau}{h^\lor}) $ is then the usual renormalisation-group flow definition for the mass-gap.}.  In this limit we find that \eqref{Zinstade} reduces to
\begin{align}
\lim_{m\to \infty}\Big[ e^{2\pi i \tau} \hat{Z}^{k=1}_{\Omega }(m;a) \Big] &\notag =-\frac{1}{\epsilon_1 \epsilon_2}\lim_{m\to \infty}\Big[ e^{2\pi i \tau} m^{2h^\lor} \biggl(\frac{\partial}{\partial x}\biggr)^{2 h^\lor-4} \Gnek(x,\epsilon\,; ia)\Big\vert_{x=0} \Big] \\
&\label{ZSYM}=-\frac{\Lambda^{2h^\lor} }{\epsilon_1\epsilon_2}\sum_{\alpha\in\Delta_L}\frac{1}{(\alpha\cdot ia)(\alpha\cdot ia+\epsilon)} \prod_{\beta\in\Delta(\alpha)}\frac{1}{\beta\cdot ia}\,,
\end{align}
thus confirming that the conjectural expression \eqref{Zinstade} correctly reduces to the formula for the one-instanton partition function of pure $\mathcal{N}=2$ SYM theory derived in \cite{Keller:2011ek} from coherent states of W-algebras.

Finally, for the purpose of computing the instanton corrections to the integrated correlator \eqref{eq:firstcorr}, we want to specialise \eqref{Zinstade} to the four-sphere case, reproduced by an $\Omega$-deformation background where $\epsilon_1=\epsilon_2=1$ and $\epsilon=\epsilon_1+\epsilon_2=2$. It is then straightforward to derive from \eqref{Zinstade}
\begin{align}
\partial_m^2 \hat{Z}^{k=1}_{S^4}(m;a)\Big\vert_{m=0}=\sum_{n=0}^{2h^\lor-4} \frac{2^{n+1}}{n!} \mathcal{E}_n\biggl(\frac{1}{2} \biggr) \biggl(\frac{\partial}{\partial x}\biggr)^n \Gnek(x,\epsilon=2\,;ia)\Big\vert_{x=0}\,. \label{eq:ZnekS4}
\end{align}
We will shortly see that the one-instanton contribution to the integrated correlator computed from this expression agrees with the predictions we will derive and which are based entirely on GNO duality and perturbative data, thus strengthening our claim that \eqref{Zinstade} is the general one-instanton Nekrasov partition function.

We now discuss separately the simply laced ADE groups and the non-simply laced groups and the consequences of the conjectural expression \eqref{eq:ZnekS4} for the one-instanton contributions to the integrated correlator \eqref{eq:firstcorr}.

\paragraph{Simply laced gauge groups (ADE)} 

Let us restrict our attention to the case where $G$ is a simply laced group, i.e. $G=SU(N)$, $SO(2N)$ and $E_{6,7,8}$.
As already mentioned, we now have $\Delta_L = \Delta$, i.e. all roots are ``long'', and since for all roots $\alpha \in \Delta$ we have $(\alpha\cdot\alpha)=2$, the set $\Delta(\alpha)$ defined in \eqref{eq:Deltaalpha} reduces to
\begin{align}
\Delta(\alpha)=\{\beta \in \Delta\,\,\vert\,\,(\alpha \cdot \beta)=1 \}\,,
\end{align}
so that the auxiliary function $\Gnek(x,\epsilon\,;a)$ can be written as 
\begin{align}
\label{GfunctionADE}
\Gnek(x,\epsilon\,;a)=\sum_{\alpha\in\Delta}\frac{1}{(\alpha\cdot a)(\alpha\cdot a+\epsilon)} \prod_{(\alpha \cdot \beta)=1}\biggl(1+\frac{x}{\beta\cdot a}\biggr)\,.
\end{align}
With these considerations, it is easy to see that in the ADE case the conjectural expression \eqref{Zinstade} is not new and it reduces to the results derived in \cite{Billo:2015pjb}.

Note that, while for $E_{6,7,8}$ there is no ADHM construction for the instanton moduli space, for the cases $G=SU(N)$, $SO(2N)$ an ADHM construction does exist and the computations obtained through equivariant localisation method can be shown \cite{Billo:2015pjb} to reproduce \eqref{Zinstade}.
However, it is important to mention that when comparing to the equivariant localisation results, \eqref{Zinstade} reproduces the correct one-instanton partition function up to $a$-independent terms. This is however not an issue for the present purpose since our goal is to evaluate the instanton contribution, $\C_G^{(k=1)}(\tau_2)$, to the integrated correlator as given in \eqref{eq:kinst}. We easily see that, thanks to the presence in \eqref{eq:kinst} of the laplacian $\Delta_{\tau}$, any $a$-independent term in the instanton partition function $\eqref{Zinstade}$ does not contribute to $\C_G^{(k=1)}(\tau_2)$.

For the cases $G=SU(N)$, $SO(2N)$, the equivariant localisation results derived in \cite{Billo:2015pjb} were used in \cite{Dorigoni:2022zcr} to compute the instanton contributions, $\C_G^{(k)}(\tau_2)$, to the integrated correlator for low instanton numbers, $k$. For this reason, we can focus our attention to the instanton corrections to the integrated correlator in the exceptional case $G=E_{6,7,8}$ for which we can only use \eqref{Zinstade} as there is no analogue ADHM counterpart.

Using the Lie algebra data collected in appendix \ref{app:Lie}, we specialise \eqref{Zinstade} to $G=E_{6,7,8}$ and compute the one-instanton partition function, $\C_G^{(1)}(\tau_2)$, from the general expression \eqref{eq:kinst}. 
A technical obstacle to computing the matrix model integral \eqref{measure} arises from the fact that these groups all have a large number of roots, thus making any manipulation of the Vandermonde determinant \eqref{eq:vdm} extremely arduous to achieve on a laptop.

We content ourselves with considering the case $G=E_6$ as we do not expect any fundamental difference with respect to the remaining simply laced exceptional cases $E_{7}$ and $E_{8}$.
The one-instanton contribution $\C_{E_6}^{(1)}(\tau_2)$ is computed in terms of the matrix model integral \eqref{eq:kinst} which we find convenient to evaluate in perturbation theory, $g_{_{Y\!M}}^2\to0$, by expanding the instanton partition function \eqref{eq:ZnekS4} for small $a_i$. 
Specialising  \eqref{eq:ZnekS4} to $G=E_6$, it is then straightforward (although computationally intense) to compute the expectation value \eqref{measure} and obtain the first two  orders in perturbation expansion for the one-instanton contribution to the $E_6$ integrated correlator 
\begin{equation}
 \C^{(1)}_{E_6}(y)=-\frac{1971567}{4194304} -\frac{2485431}{4194304}y^{-1}
 + O(y^{-2})\,, \label{eq:E6k1}
\end{equation}
where as usual we define $y:=\pi\tau_2=\frac{4\pi^2}{g_{Y\!M}^2}$.
We will retrieve this result later in section \ref{sec:ansatz} starting from the conjectural exact lattice-sum expression \eqref{gsun}.

\paragraph{Non-simply laced gauge groups}

In this case, the one-instanton partition function, $ \hat{Z}^{k=1}_{\Omega }(m;a)$, in the $\Omega$-deformation background is in general not known and we have to conjecture the validity of our proposed \eqref{Zinstade}. Let us begin by making some considerations.

First of all, as already remarked in the flat-space limit, $\epsilon_1,\epsilon_2\to0$, the conjectural expression simplifies dramatically to \eqref{eq:ZnekflatNSL} thus reproducing the results of \cite{Billo:2015jyt} for the one-istanton partition function in the $\mathcal{N}=2^*$ SYM.
Secondly as discussed above, the decoupling limit, $m\to\infty$, of \eqref{Zinstade} reduces to \eqref{ZSYM} thus matching the result derived in \cite{Keller:2011ek} for the one-instanton partition function in a general $\Omega$-deformation background for pure $\mathcal{N}=2$ SYM.
Moreover, in the non-simply laced cases $G=SO(2N+1)$, $USp(2N)$ an ADHM construction for the instanton moduli space is possible and we can compare \eqref{Zinstade} against the results obtained through equivariant localisation method \cite{Billo:2015jyt} finding perfect agreement (again modulo $a$-independent terms which do not affect our integrated correlator).  

We stress that in our conjectural expression \eqref{Zinstade}, the only difference between the ADE case for which this formula was proven in \cite{Billo:2015pjb} and the non-simply laced cases is that in the auxiliary function, $\Gnek(x,\epsilon\,;a)$, defined in \eqref{Gfunction} the sum runs over all roots, $\alpha\in \Delta$, in the ADE case and only over \textit{long} roots, $\alpha\in \Delta_L$, in the non-simply laced case. 

Again, for the classical cases $G=SO(2N+1)$, $USp(2N)$ the instanton contributions, $\C_G^{(k)}(\tau_2)$, to the integrated correlator have been computed in  \cite{Dorigoni:2022zcr}  for low instanton numbers, $k$, using the equivariant localisation results derived in \cite{Billo:2015jyt}. Here we focus on the remaining exceptional cases $G=F_4$ and $G_2$ for which we can only rely on \eqref{Zinstade} as there is no analogue ADHM counterpart.

Just as we did before, the one-instanton contribution $\C_{F_4}^{(k=1)}(\tau_2)$ is computed in terms of the matrix model integral \eqref{eq:kinst}. Specialising  \eqref{eq:ZnekS4} to $G=F_4$ does not produce a particularly pretty expression which, as such, will not be presented here. However, using the definition \eqref{measure} for the matrix model expectation value, we use \eqref{eq:ZnekS4} to arrive at
\begin{align}
\C^{(1)}_{F_4}(y)=&\label{eq:F4instEx} \frac{13 \,y^{\threeh}}{55050240}\Big[-4\sqrt{y}\, q_1(8y)  +  \sqrt{\pi } e^{4 y} \text{erfc}\left(2 \sqrt{y}\right)q_2(8y) \Big]\,,
\end{align}
where $\text{erfc}(z):=\frac{2}{\sqrt{\pi}}\int_z^\infty e^{-t^2}dt$ is the complementary error function, while $q_{1}(y)$ and $q_2(y)$ are the two particularly unilluminating polynomials
\begin{align}
q_1(y) &:=\notag 17 y^7+2863 y^6+176175 y^5+5110965 y^4+73337163 y^3+494651997 y^2\\
&\phantom{=}+1354678605 y+1026765495\,,\\
q_2(y) &:=\notag 17 y^8+2880 y^7+179004 y^6+5281584 y^5+78123150 y^4+559329120 y^3\\
&\phantom{=}+1742611500 y^2+1850189040 y+294440265\,.
\end{align}
For later purposes we find it convenient to expand \eqref{eq:F4instEx} in perturbation theory, i.e. for $y\gg1$,  
\begin{equation} 
\C^{(1)}_{F_4}(y)=-\frac{117}{512}+\frac{4797}{4096}y^{-1}-\frac{78975}{32768}y^{-2}+\frac{552825}{131072}y^{-3}+O(y^{-4})\,.\label{eq:F4k1}
\end{equation}

Finally for $G_2$ we can write the one-instanton partition function \eqref{eq:ZnekS4} in the compact form
\begin{align}
\partial_m^2 \hat{Z}^{k=1}_{G_2}(m,a_i)\big\vert_{m=0}=-\frac{18(3a_1^2+3a_2^2+2)^2}{(3a_2^2+2)[ 9(3a_1^2-a_2^2)^2+48(3a_1^2+a_2^2)+64]}\,,
\end{align}
which allows us to evaluate explicitly the matrix model integral \eqref{eq:kinst},
\begin{equation}
\C_{G_2}^{(1)}(y)=\frac{21}{128}y^{\threeh}\Big\lbrace -4\sqrt{y}\,[3 (8y)^2+62 (8y)+175]+\sqrt{\pi}e^{4y}\,\text{erfc}(2\sqrt{y})[3 (8y)^3+65 (8y)^2+231 (8y)+81] \Big\rbrace\,.\label{eq:G2instEx}
\end{equation}

Once again it will result useful to expand \eqref{eq:G2instEx} in perturbation theory
\begin{equation} 
\C^{(1)}_{G_2}(y)=-\frac{63}{128}+\frac{63}{64}y^{-1}-\frac{4725}{4096 }y^{-2}+\frac{19845}{16384}y^{-3}+O(y^{-4})\,.\label{eq:G2k1}
\end{equation}

In the next section \ref{sec:ansatz} we will show that these results can be reproduced starting from the completely independent exact lattice-sum expression \eqref{gsun}, thus strengthening the claim that the conjectural one-instanton Nekrasov partition function \eqref{Zinstade} is indeed valid for all simple groups.

Motivated by our perturbative results discussed in section \ref{pertexpn}, it would be extremely interesting to obtain a universal expression (if any) also for the one-instanton Nekrasov partition function or directly for the one-instanton (and higher) contribution to the integrated correlator. In this regard a universal formula for the one-instanton partition function of pure $\mathcal{N}=2$ SYM theory \eqref{ZSYM} was derived in \cite{Mkrtchyan:2016ehw} for the special case where the vacuum expectation value of the scalar field, $a$, is restricted to the Weyl line. 
To us it is not obvious how to extend the analysis of \cite{Mkrtchyan:2016ehw} to generic values of the scalar field $a$ or how to promote these results to $\cN=2^*$ by incorporating a massive adjoint hypermultiplet.
It may be possible that such a universal expression for the general Nekrasov partition function does not exist, while a universal expression can only be obtained after we have performed the matrix model integral \eqref{measure} for Nekrasov partition function, i.e. only for the integrated correlator instanton sectors, $\C^{(k)}_G(\tau_2)$.

\section{The duality covariant ansatz}
\label{sec:ansatz}

In this section we motivate the validity of the conjectural expression for $\C_{G}(\tau)$ written in terms of the Montonen-Olive duality covariant lattice sum \eqref{gsun} and rewritten here for convenience
\begin{equation}
\C_{G}(\tau)  = \sum_{(m,n)\in \Z^2} \int_0^\infty \Big[e^{- t \pi \frac{|m+n\tau|^2}{\tau_2}} B_{\mathfrak{g}}(t) +e^{- t \pi \frac{|m+ n \mathfrak{n}_{\mathfrak{g} }\tau|^2}{\mathfrak{n}_{\mathfrak{g}}  \tau_2}}  B_{^L \mathfrak{g}}(t) \Big] dt\,. \label{gsun2}
\end{equation}
Given a simple gauge group $G$ we denote by $\mathfrak{g}$ the associated Lie algebra and by $^L\mathfrak{g}$ its GNO (or Langlands) dual algebra which will be  reviewed shortly.

The integrand, $B_\mathfrak{g}(t)$, is a rational function of the form
 \begin{equation}\label{eq:Bgeneral}
B_{\mathfrak{g}} (t)= \frac{\mathcal{Q}_{\mathfrak{g}}(t) }{ (t+1)^{2 h^\lor+1} }\,,
 \end{equation}
 where $h^\lor$ is the dual Coxeter number of the Lie algebra $\mathfrak{g}$,  while $\mathcal{Q}_{\mathfrak{g}}(t)$ is a degree $2 h^\lor-1$ polynomial without constant term.
 The functions $B_{\mathfrak{g}} (t)$ satisfy the following identities 
\begin{align}
&\label{eq:Bident1} \int_0^\infty B_\mathfrak{g}(t) dt = -b_\mathfrak{g}(0)\,, \qquad \int_0^\infty B_\mathfrak{g}(t)\frac{dt}{\sqrt{t}} =0\,,\\ 
&\label{eq:Bident2} B_{\mathfrak{g}}(t) = t^{-1} B_{\mathfrak{g}}(t^{-1})\,,
\end{align}
where the last identity is a consequence of the ``palindromic'' property $\mathcal{Q}_{\mathfrak{g}}(t)=t^{2 h^\lor } \mathcal{Q}_{\mathfrak{g}}(t^{-1})$.

The Lie algebra dependent numbers $b_\mathfrak{g}(0)$ where determined in \cite{Dorigoni:2022zcr} for all classical gauge groups,
\begin{align}
b_{\mathfrak{su}_N}(0) &\notag = -\frac{N(N-1)}{16}\,, \qquad b_{\mathfrak{so}_{2N}}(0) = -\frac{N(N-1)}{8}\,,\\
b_{\mathfrak{so}_{2N+1}}(0)& = -\frac{N(N-1)}{4}\,, \qquad b_{\mathfrak{sp}_{2N}}(0)  = -\frac{N}{4}\,,
\end{align}
while promptly we will show that
\begin{equation}
b_{\mathfrak{g}_2}(0) = -\frac{3}{4}\,,\qquad b_{\mathfrak{f}_4}(0) = -3\,,\qquad b_{\mathfrak{e}_6}(0) = -\frac{9}{2}\,.\label{eq:Bint}
\end{equation}

Concretely, for $SU(N)$ we have (remembering that $2h^\lor =2N$ for $SU(N)$)
 \ie \label{eq:BSUN}
 B_{\mathfrak{su}_{N}}(t)= \frac{\mathcal{Q}_{\mathfrak{su}_{N}}(t)}{(t+1)^{2N+1}}\, , 
 \fe  with 
\begin{align}
 \mathcal{Q}_{\mathfrak{su}_{N}}(t) &\nn = -\frac{1}{8}N(N-1)(1-t)^{N-1}(1+t)^{N+1}\\
 &\left\lbrace [3+(8N+3t-6)t] P_N^{(1,-2)}\Big(\frac{1+t^2}{1-t^2}\Big) +\frac{1}{1+t} (3t^2-8 N t-3 )P_{N}^{(1,-1)}\Big(\frac{1+t^2}{1-t^2}\Big)\right\rbrace\,,
 \label{eq:Bndef}
 \end{align}
 expressed in terms of Jacobi polynomials $P_n^{(a,b)}(x)$. 
This expression was first conjectured in \cite{Dorigoni:2021bvj,Dorigoni:2021guq} and then proven in \cite{Dorigoni:2022cua} using matrix-models methods. Similarly, in \cite{Dorigoni:2022zcr} analogous formulae were derived for $ B_{\mathfrak{g}}(t)$ in the case of general classical gauge groups $\mathfrak{g} = \mathfrak{so}_n$ (with $n=2N, 2N+1$) and $\mathfrak{g}= {\mathfrak{sp}_{2N}}$ subsequently expressed in \cite{Dorigoni:2022cua} via generating series.

In \cite{Dorigoni:2021bvj,Dorigoni:2021guq,Dorigoni:2022zcr} the authors arrived at \eqref{gsun2} by combining supersymmetric localisation calculations, as discussed in section \ref{intgencorr}, with constraints coming from Montonen--Olive duality. Before we review and extend their analysis, let us present two different decompositions for \eqref{gsun2}.

A first decomposition can be obtained by substituting the Taylor series expansion, 
\begin{equation}
B_\mathfrak{g}(t) = \sum_{s=2}^\infty b_{\mathfrak{g}}(s) \,\frac{t^{s-1}}{\Gamma(s)}\,,\label{eq:Btaylor}
\end{equation}
in \eqref{gsun2} and then performing the $t$-integral term by term, arriving at
\begin{equation}
\C_{G}(\tau) = -\Big[ b_{\mathfrak{g}}(0)+b_{^L\mathfrak{g}}(0)\Big] + \sum_{s=2}^\infty \Big[ b_{\mathfrak{g}}(s) \,{\rm E}(s;\tau) +b_{^L\mathfrak{g}}(s) \,{\rm E}(s;n_\mathfrak{g}\tau)\Big]\,.\label{eq:Eisenexp}
\end{equation}
Here ${\rm E}(s;\tau)$ denotes the non-holomorphic (as a function of $\tau$) Eisenstein series,
\begin{equation}
{\rm E}(s;\tau):= \sum_{(m,n)\neq (0,0)} \frac{(\tau_2/\pi)^s}{|m+n\tau|^{2s}} = \sum_{(m,n)\neq (0,0)} \int_0^\infty e^{- t \pi \frac{|m+n\tau|^2}{\tau_2}} \frac{t^{s-1}}{\Gamma(s)}\, dt\,,
\label{eq:Eisen}
\end{equation}
which is an invariant function under the action of the modular group ${ SL}(2,\Z)$,
\begin{equation}
{\rm E}(s;\gamma\cdot\tau)={\rm E}(s;\tau)\,,
\end{equation}
where as usual we defined
\begin{equation}
 \gamma\cdot \tau:= \frac{a\tau+b}{c\tau+d}\,,\qquad \gamma = \left(\begin{matrix} a& b \\ c & d \end{matrix}\right)\in { SL}(2,\Z)\,.
\end{equation}
In~\cite{Collier:2022emf} it was pointed out that for $G= SU(N)$, the formal expansion~\eqref{eq:Eisen} can be rewritten as a conventional and convergent spectral decomposition for $\C_{SU(N)}$. This formal decomposition \eqref{eq:Eisenexp}, will be very useful when we come back to discussing the covariance of \eqref{gsun2} under GNO duality.

Another useful way of decomposing the lattice sum \eqref{gsun2} is via its Fourier mode expansion with respect to $\tau_1 = \frac{\theta}{2\pi}$,
\begin{align*}
\C_{G} (\tau) = \C_{G}^{(0)} (\tau_2) + \sum_{k=1}^\infty  \left(e^{2\pi i k \tau}\, \C_{G}^{(k)}(\tau_2) +e^{-2\pi i k \bar\tau}\, \C_{G}^{(-k)}(\tau_2) \right) \, .
\end{align*}
Following \cite{Dorigoni:2021guq}, we perform a Poisson resummation of the lattice sum \eqref{gsun2} over the summation variable $m\to\hat{m}$ and then use the identities \eqref{eq:Bident1}-\eqref{eq:Bident2} to arrive at the Fourier zero-mode formula
\begin{align}
\C_{G}^{(0)} (\tau_2)  &=\label{eq:0mode} 4\sum_{n>0} \int_0^\infty \Big[ \exp(-t\pi n^2  \tau_2)  \sqrt{\tau_2}\,B_{\mathfrak{g}}(t)+  \exp(-t\pi n^2 \mathfrak{n}_{\mathfrak{g}} \tau_2) \,\sqrt{ \mathfrak{n}_{\mathfrak{g}}\tau_2} \,B_{^L \mathfrak{g}}(t) \Big]  \frac{dt}{\sqrt{t}}\\
&\notag \sim \sum_{s=2}^\infty \frac{4 \Gamma(s-\half)}{\sqrt{\pi} \Gamma(s)} \Big[b_{\mathfrak{g}}(s)  + (\mathfrak{n}_\mathfrak{g})^{1-s}b_{^L\mathfrak{g}}(s) \Big] \zeta_{2s-1}y^{1-s} \,,
\end{align}
where $y:=\pi\tau_2$.
In a similar fashion, we obtain the $k$-instanton sector expression (with $k\neq0$)
\begin{align}
 \C_{G}^{(k)} (\tau_2) = &\label{eqkinst}\sum_{\hat{m}n = k}\int_0^\infty \exp\Big[ - \Big( \frac{|\hat{m}|}{\sqrt{t}} - |n| \sqrt{t}\Big)^2\pi\tau_2\Big]\, \sqrt{\tau_2}\,B_{\mathfrak{g}}(t)\,\frac{dt}{\sqrt{t}}\\
 &\notag +\sum_{\hat{m}n = \frac{k}{ \mathfrak{n}_\mathfrak{g} } }\int_0^\infty \exp\Big[ -\Big( \frac{|\hat{m}|}{\sqrt{t}} - |n| \mathfrak{n}_\mathfrak{g} \sqrt{t}\Big)^2 \pi\tau_2 \Big] \,\sqrt{\mathfrak{n}_\mathfrak{g}\tau_2}\,B_{\!^L\mathfrak{g}}(t)\,\frac{dt}{\sqrt{t}}\,.
\end{align}
Note that since $\hat{m},n\in \mathbb{Z}$, in the case of non-simply laced groups where $\mathfrak{n}_\mathfrak{g}\neq1$, the second contribution to the $k$-instanton sector is non-vanishing only for instanton numbers which are a multiple of $\mathfrak{n}_\mathfrak{g}$, i.e. only when $k \equiv 0\,(\rm{mod}\, \mathfrak{n}_\mathfrak{g})$. We will come back to \eqref{eq:0mode}-\eqref{eqkinst} when discussing the particular cases of exceptional gauge groups $G$.

We now briefly review how S-duality (Montonen--Olive duality) acts on $\mathcal{N}=4$ SYM with general gauge group and the crucial r\^ole that this property plays in constructing the lattice sum ansatz \eqref{gsun2}.

\subsection{Goddard--Nuyts--Olive duality}
\label{secGNO}

In a now classic paper by Goddard, Nuyts, and Olive~\cite{Goddard:1976qe} it was shown that in gauge theories electric charges are associated with the weight lattice of the gauge group $G$, while magnetic charges take values in the lattice of a dual group denoted by $^L{G}$. 
From the mathematical point of view, GNO duality is closely connected to the Langlands programme \cite{Kapustin:2006pk} hence we use the superscript on $^LG$ to denote the Langlands, or equivalently GNO, dual group. 
The GNO dual groups corresponding to all simple Lie groups are listed in Table~\ref{table1}-(a). 

Building on this, Montonen and Olive~\cite{Montonen:1977sn} conjectured the existence of an electro-magnetic duality between a gauge theory with gauge group $G$ and coupling constant $g_{_{Y\!M}}$, and a different theory featuring $^LG$ as its gauge group and with coupling $^Lg_{{_{Y\!M}}}=4\pi/g{_{_{Y\!M}}}$. 
Witten~\cite{Witten:1978mh} then realised that for such a duality to hold the theory under consideration had to be supersymmetry and Osborn~\cite{Osborn:1979tq} finally argued that $\cN=4$ SYM should realise GNO duality. 

As already stressed, since the integrated correlators studied in the present work only involve local operators, they are not sensitive to global properties of the groups $G$ and $^LG$. For this reason we can discuss the effects of GNO duality on the integrated correlators \eqref{eq:firstcorr} directly at the level of Lie algebras as shown by the labels in Table~\ref{table1}-(b) .  

\begin{table}[tp]
\centering
\begin{tabular}{ cc }   
\begin{tabular}{|c|c|}
 \hline
$G_N \phantom{\Big\vert}$&$\!\!^L{G_N}$\\ [0.3ex] 
 \hline\hline
$U(N)$  &  $U(N)$\\
$SU(N)$&$PSU(N) = SU(N)/\Z_N$\\
$Spin(2N)$ &$SO(2N)/\Z_2$\\
$Sp(N) = USp(2N)$  &   $SO(2N+1)$\\
$Spin(2N+1)$ & $Sp(N)/\Z_2= USp(2N)/\Z_2$\\
$E_{r=6,7,8}$ & $ E_{r}/\Z_{9-r}$\\
$F_4$ & $ F_4$\\
$G_2$ & $G_2$  \\[1ex] 
 \hline
 \end{tabular}
 &  \qquad \qquad
 \begin{tabular}{|c| c|} 
 \hline
 $\phantom{\Big\vert}\mathfrak{g}_N$&$\!\!^L{\mathfrak{g}_N}$ \\ [0.3ex] 
 \hline\hline
$\mathfrak{su}_N$&$\mathfrak{su}_N$\\
$\mathfrak{so}_{2N}$ &$\mathfrak{so}_{2N}$\\
$\mathfrak{sp}_{2N}$ & $\mathfrak{so}_{2N+1}$\\
$\mathfrak{so}_{2N+1}$  &   $\mathfrak{sp}_{2N}$\\
$\mathfrak{e}_{6,7,8}$ & $\mathfrak{e}_{6,7,8}$\\
$\mathfrak{f}_4$ & $\mathfrak{f}_4$\\
$\mathfrak{g}_2$ & $\mathfrak{g}_2$ \\ [1ex] 
 \hline
 \end{tabular}
 \vspace{0.2cm}
 
 \\ 
 (a)~Langlands/GNO relation between   & \qquad\qquad (b)~ Duality relations \\
classical Lie groups and their dual groups. & \qquad\qquad of relevance to this paper.
\end{tabular}
\caption{  } \label{table1}
\end{table}

Let us consider $\mathcal{N}=4$ SYM with complexified coupling constant 
$$
\tau= \tau_1+i\tau_2 := \frac{\theta}{2\pi} + i \frac{4\pi}{g_{_{Y\!M}}^2}\,,
$$
and general gauge group, $G$.
As discussed in~\cite{Girardello:1995gf,Dorey:1996hx},  the extended duality group of the theory is obtained by combining periodicity in the topological angle,~$\theta$,  with electro-magnetic S-duality which inverts the coupling constant while changing the theory from having gauge group $G$ to $^LG$. These duality transformations, denoted by~$T$ and~$\hat S$, are defined by 
\bea
\label{sduality}
&& T\, :\, (G,\tau) \to (\,G, \, \tau+1)\,, \nn\\
&&\hat S\, :\, (G,\tau) \to (^LG, -\frac{1}{\mathfrak{n}_\mathfrak{g} \tau})\,,
\eea
where $\mathfrak{n}_\mathfrak{g}$ denotes the ratio between the length square of the long and short roots of $\mathfrak{g}$.
Following standard conventions summarised in appendix \ref{app:Lie}, we have
\begin{align}
\mathfrak{n}_\mathfrak{g} &\notag = 1\,, \qquad \mathfrak{g} = \{\mathfrak{su}_N,\,\mathfrak{so}_{2N},\, \mathfrak{e}_6,\,\mathfrak{e}_7,\, \mathfrak{e}_8\}\,,\\
\mathfrak{n}_\mathfrak{g}&\label{eq:ng}=2\,,\qquad \mathfrak{g}= \{\mathfrak{so}_{2N+1},\,\mathfrak{sp}_{2N},\,\mathfrak{f}_4\}\,,\\
\mathfrak{n}_\mathfrak{g}&\notag = 3 \,,\qquad\mathfrak{g} = \mathfrak{g}_2\,.
\end{align}

Let us now discuss what the consequences of GNO duality are for the integrated correlator \eqref{eq:firstcorr} and how such constraints can be combined with the supersymmetric localisation results discussed in section \ref{intgencorr} to arrive at the lattice sum representation \eqref{gsun2}. As one can easily imagine from equation \eqref{eq:ng} and Table \ref{table1}-(b), simply laced gauge groups,~i.e.~$G=SU(N)$, $SO(2N)$, $E_{6,7,8}$, and non-simply laced groups,~i.e.~$G=SO(2N+1)$, $USp(2N)$, $F_4$, $G_2$, have fundamental differences under GNO duality. For this reason we will discuss the two cases separately.

\subsection{Simply laced gauge groups (ADE)}

In the simply laced cases, i.e. for gauge groups of ADE type, we see from \eqref{eq:ng} that  $\mathfrak{n}_\mathfrak{g}=1$, hence 
$$\hat S \equiv S: \tau\to - \frac{1}{\tau}\,,$$
reduces to usual S-duality. Furthermore, we can see from Table \ref{table1}-(b) that the corresponding Lie algebras are Langlands self-dual, i.e. $\mathfrak{g}=\!\!\,^L\mathfrak{g}$, so that for ADE gauge groups $S$ and $T$ are genuine symmetries of the theory (for local observables). When combined, $S$ and $T$ generate the discrete self-duality group $SL(2,\Z)$, which acts in the usual way on the complexified coupling constant,
\begin{equation}
 \tau  \underset {SL(2,\Z)} \to \gamma\cdot \tau= \frac{a\tau+b}{c\tau+d}\,,\qquad \gamma = \left(\begin{matrix} a& b \\ c & d \end{matrix}\right)\in { SL}(2,\Z)\,. \label{sl2t}
\end{equation}
Since for all ADE algebras we have $\mathfrak{g}=\!\!\,^L\mathfrak{g}$, the corresponding Borel transforms must satisfy
\begin{equation}
B_{\mathfrak{g}}(t) = B_{^L\mathfrak{g}}(t)\,,
\end{equation}
hence the lattice sum formula \eqref{gsun2} reduces to
\begin{align}
\C_{G}(\tau)  &= 2\sum_{(m,n)\in \Z^2} \int_0^\infty e^{- t \pi \frac{|m+n\tau|^2}{\tau_2}} B_{\mathfrak{g}}(t) \,dt\,, \label{gADE}\\
&\notag = -2 b_{\mathfrak{g}}(0)+ 2\sum_{s=2}^\infty  b_{\mathfrak{g}}(s) \,{\rm E}(s;\tau) 
\end{align}
which is manifestly invariant under the expected GNO duality group, $SL(2,\Z)$.

In \cite{Dorigoni:2021bvj,Dorigoni:2021guq,Dorigoni:2022zcr}\footnote{Note that these references use a slightly different convention. In particular when comparing with the present case the Borel transform, $B_G(t)$, defined in \cite{Dorigoni:2022zcr} for $G=SU(N),SO(2N)$, we have $B_G(t) = B_{\mathfrak{g}}(t) +B_{^L\mathfrak{g}}(t) = 2 B_{\mathfrak{g}}(t) $.}, the authors used supersymmetric localisation, and in particular the general results of \cite{Alday:2021vfb}, to compute the perturbative expansion as $g_{_{Y\!M}}\to0$ of the integrated correlator \eqref{eq:firstcorr} for $G=SU(N)$ and $SO(2N)$, as discussed in section \ref{intgencorr}. Assuming the validity of this conjectural expression \eqref{gADE}, it is possible to use such perturbative data to reconstruct $B_{\mathfrak{g}}(t)$ from the Fourier zero-mode expansion \eqref{eq:0mode}.
Finally as an independent check, the $k$-instanton sector was computed in two different ways: either as reviewed in section \ref{sec:yminst} directly from Nekrasov instanton partition function or from the Fourier mode decomposition for the candidate expression \eqref{eqkinst}. The two independent calculations yield the same result thus confirming the validity of \eqref{gADE} for $G=SU(N)$ and $SO(2N)$, which was subsequently proven by other means in \cite{Dorigoni:2022cua}.

We repeat the same method here for the exceptional series $G=E_{6}$, $E_7$, and $E_8$. Firstly, we can use the supersymmetric localisation formula \eqref{eq:ZGpert}, or more efficiently the universal expansion \eqref{pertexp}, to compute the perturbative expansion as $g_{_{Y\!M}}\to0$ of the exceptional integrated correlators.
Then from the equivalent Fourier zero-mode expansion \eqref{eq:0mode} we reconstruct $B_{\mathfrak{g}}(t)$ which in the case of $\mathfrak{g}= \mathfrak{e}_6$ takes the manifest GNO invariant form 
\begin{align}
\C_{E_6}(\tau) & =2 \sum_{(m,n)\in \Z^2} \int_0^\infty e^{- t \pi \frac{|m+n\tau|^2}{\tau_2}}  B_{\mathfrak{e}_6}(t)\, dt\\
&\nn= 9+ 2\sum_{s=2}^\infty b_{\mathfrak{e}_6}(s)\, { E}(s; \tau)\,,
\end{align}
with (remembering that $2h^\lor=24$ for $\mathfrak{e}_6$)
\begin{equation}
B_{\mathfrak{e}_6}(t) =\frac{ {\mathcal{Q}}_{\mathfrak{e}_6}(t) }{(t+1)^{25}}\,,\label{eq:BE6}
\end{equation} 
and the palindromic polynomial 
\begin{align}
{\mathcal{Q}}_{\mathfrak{e}_6}(t) &:= 351\big[ p_{\mathfrak{e}_6}(t)+ t^{24}p_{\mathfrak{e}_6}(t^{-1}) \big]= t^{24} {\mathcal{Q}}_{\mathfrak{e}_6}(t^{-1})\,,\\
p_{\mathfrak{e}_6}(t) &\notag:= t-25 t^2+226 t^3 -1390 t^4+5110 t^5-\frac{30227 t^6}{2}+\frac{67925 t^7}{2}-67036 t^8+105190 t^9\\
&\notag\phantom{:=}-150220 t^{10}+179578 t^{11}-95900 t^{12}\,.
\end{align}
Given the explicit formula \eqref{eq:BE6} for $B_{\mathfrak{e}_6}(t) $, we can easily check that it satisfies the identities \eqref{eq:Bident1}-\eqref{eq:Bident2}-\eqref{eq:Bint}.

From this conjectural expression for $B_{\mathfrak{e}_6}(t) $, we can compute using \eqref{eqkinst} the general $k$-instanton sector for the integrated correlator with gauge group $G=E_6$. In particular for $k=1$ we easily find
\begin{align}
\C^{(1)}_{E_6}(y)=&\label{eq:E6instEx} \frac{13 \,y^{\threeh}}{216466351718400}\Big[4\sqrt{y}\, q_1(8y)  -  \sqrt{\pi } e^{4 y} \text{erfc}\left(2 \sqrt{y}\right)q_2(8y) \Big]\,,
\end{align}
where the two polynomials $q_{1}(y)$, $q_{2}(y)$ are given by
\begin{align}
q_1(y) &:=\notag 7429 y^{10}+2386970 y^9+308720493 y^8+21069931512 y^7+832047379170 y^6\\
&\phantom{=}+19598615078220 y^5+274167552296250 y^4+2203639597269720 y^3\\
&\phantom{=} \notag +9483953753164905 y^2+18934598296463850 y+12157131999069225\,,\\
q_2(y) &:=\notag 7429 y^{11}+2394399 y^{10}+311092605 y^9+21373952355 y^8+852523189650 y^7\\
&\phantom{=}\!\!\!\!\!\!\!\!\!+20391438408390 y^6+292291551180090 y^5+2445570512724150 y^4\\
&\phantom{=} \!\!\!\!\!\!\!\!\! \notag +11286279339248025 y^3+25748332503460875 y^2+22782924235560825 y+3428445324655575\,.
\end{align}
We then expand \eqref{eq:E6instEx} for $y\gg1$ to arrive at 
\begin{equation}
 \C^{(k=1)}_{E_6}(y)=-\frac{1971567}{4194304} -\frac{2485431}{4194304 }y^{-1}
 +\frac{148983705}{134217728 }y^{-2}-\frac{1751460165}{536870912 }y^{-3}
 + O\Big(y^{-4}\Big)\,, \label{eq:E6k1ex}
\end{equation}
nicely matching (and extending) the result \eqref{eq:E6k1} previously found from the independent supersymmetric localisation calculation starting from the conjectural form \eqref{eq:ZnekS4} for exceptional Nekrasov partition function. 

For the remaining two simply laced cases, $G=E_{7}$ and $E_8$, the story is analogous albeit extremely more demanding from a computational point of view. 
If the conjectural expression \eqref{eq:Bgeneral} for $B_\mathfrak{g}(t)$ is correct, the palindromic polynomial, $\mathcal{Q}_{\mathfrak{g}}(t)$, is specified by its first $h^\lor$ coefficients, with $h^\lor$ the dual Coxeter number of $\mathfrak{g}$. We deduce that to determine uniquely $\mathcal{Q}_{\mathfrak{e}_7}(t)$, and $\mathcal{Q}_{\mathfrak{e}_8}(t)$, we need to compute $18$, respectively $30$, orders in perturbation theory. 
Unfortunately, if we try to attack this computation using the supersymmetric localisation formula \eqref{eq:ZGpert} we immediately encounter a technical issue.

At each order in perturbation theory, we need evaluating very standard matrix model integrals $\langle (\alpha\cdot a)^{2k}\rangle_G$, with $k\in \mathbb{N}$, simple moments of a multi-variable gaussian distribution~\eqref{eq:Sclass} with the notable feature that the measure factor does contain the Vandermonde determinant \eqref{eq:vdm} as well. Given that Vandermonde determinant contains $63$ positive roots for $E_7$ and $120$ positive roots for $E_8$, we have not managed to evaluate this matrix model integral with any computer algebra program running on our laptops. Since no fundamental change is expected between $E_6$ and the missing $E_7$ and $E_8$ cases, we content ourselves with the present analysis.

In \cite{Alday:2021vfb} following the earlier work \cite{Chester:2019pvm}, an alternative and more efficient approach was used for determining exactly the perturbative sector of all integrated correlators with classical gauge groups in terms of a simple integral transform of generalised Laguerre polynomials. In particular all Vandermonde determinants for classical groups can be simplified dramatically by using Hermite orthogonal polynomials. We are not aware of similar results for exceptional groups but such methods could potentially lead to a more efficient way of computing arbitrarily higher orders in perturbation theory for $E_7$ and $E_8$.

\subsection{Non-simply laced gauge groups}

In the non-simply laced cases we need to distinguish between the classical groups,~i.e. $G=SO(2N)$ and $USp(2N)$, and the two exceptional cases $G=F_4$ and $G_2$.

The classical cases were discussed in \cite{Dorigoni:2022zcr} so that here we only summarise their results.
From Table \ref{table1}-(b), we see that $\mathfrak{sp}_{2N} =\!^L\mathfrak{so}_{2N+1}$ and similarly $\mathfrak{so}_{2N+1} =\!^L\mathfrak{sp}_{2N}$. Furthermore from \eqref{eq:ng} we see that in both cases $\mathfrak{n}_\mathfrak{g}=2$, so that the $\hat S$ transformation defined in \eqref{sduality} maps the theory with gauge group $USp(2N)$  into the theory with gauge group $SO(2N+1)$ and a modified coupling constant.  
From equation \eqref{gsun2} it is rather easy to see that the exact integrated correlators \eqref{eq:firstcorr} transform covariantly under GNO duality,~i.e.
\begin{equation}
 \cC_{SO(2N+1)} (\tau)=\cC_{USp(2N)}\Big({-}\frac{1}{2\tau}\Big) \,,\qquad \cC_{USp(2N)} (\tau)=\cC_{SO(2N+1)}\Big({-}\frac{1}{2\tau}\Big) \,,
\end{equation}
 consequence of
\begin{equation}
B_{\mathfrak{so}(2N+1)}(t) = B_{^L\mathfrak{sp}(2N)}(t) \,,\qquad\qquad\quad   B_{\mathfrak{sp}(2N)}(t)=B_{^L\mathfrak{so}(2N+1)}(t)\,.
\end{equation}

We note that in these cases, $\hat S$ (which is not a symmetry) acts on the coupling, $\tau$, as an ${ SL}(2,\RR)$ transformation that is not in ${ SL}(2,\Z)$. We can however consider the transformations $\hat S T\hat S$ and $T$ which are both symmetries,~i.e.~they do not change $G$, and when combined they generate a $\Gamma_0(2)$ subgroup of $SL(2,\Z)$, where we remind the reader
\begin{equation}
\Gamma_0(r) := \left\lbrace \gamma\in {\rm SL}(2,\Z)\,\Big\vert\,\gamma = \left( \begin{matrix} a & b \\ c & d \end{matrix}\right)\,{\rm{with}}\,\, c\equiv 0\,(\rm{mod}\,r) \right\rbrace \,.
\end{equation}
 In other words $\Gamma_0(2)$ is the self-duality group that leaves the integrated correlators $\C_{G}$ invariant for $G=SO(2N+1)$ and $G=USp(2N)$:
 \begin{equation}
\cC_{SO(2N+1)} (\gamma\cdot \tau) =  \cC_{SO(2N+1)} (\tau)\,,\qquad   \cC_{USp(2N)} (\gamma\cdot \tau) =  \cC_{USp(2N)} (\tau) \,,\qquad \forall \,\gamma\in \Gamma_0(2)\,.
 \end{equation}
This symmetry is manifest both from the lattice sum \eqref{gsun2} when $\mathfrak{n}_\mathfrak{g}=2$, or from the expansion in Eisenstein series \eqref{eq:Eisenexp} where clearly ${\rm E}(s;2\tau)$ is invariant under $\Gamma_0(2)$.

Once again,~\cite{Dorigoni:2022zcr} exploited the supersymmetric localisation perturbative results of~\cite{Alday:2021vfb}, to reconstruct $B_{\mathfrak{g}}(t)$ from the Fourier zero-mode expansion~\eqref{eq:0mode} for $\mathfrak{g}= \mathfrak{so}_{2N+1}$ and $\mathfrak{sp}_{2N}$. The validity of \eqref{gsun2} was then checked at the non-perturbative level by comparing the $k$-instanton sector computed from Nekrasov instanton partition function against the Fourier mode decomposition \eqref{eq:kinst}. The conjectural lattice sums expressions for $\mathfrak{g}= \mathfrak{so}_{2N+1}$ and $\mathfrak{sp}_{2N}$ were subsequently proven by other means in \cite{Dorigoni:2022cua}.

We now move to the novel exceptional, non-simply laced cases $G=F_4$ and $G_2$. There are a number of distinctive features involved in S-duality for gauge theories with exceptional groups \cite{Argyres:2006qr,Kapustin:2006pk}. 
First of all, we notice from Table \ref{table1}-(a) that both algebras are GNO self-dual,~i.e. $^L\mathfrak{f}_4=\mathfrak{f}_4$ as well as $^L\mathfrak{g}_2=\mathfrak{g}_2$, hence from \eqref{sduality} we see that the transformation $\hat{S}$ is then again a symmetry of the integrated correlators. However, both algebras have root systems  containing long \textit{and} short roots so that we must have $\mathfrak{n}_\mathfrak{g}\neq 1$ as presented in \eqref{eq:ng}. These observations have striking implications.

First of all as just discussed, the symmetry transformations $\hat S T\hat S$ and $T$ generate the congruence subgroup $\Gamma_0(\mathfrak{n}_\mathfrak{g})$. 
Furthermore given that $F_4$ and $G_2$ are GNO self-dual, the corresponding Borel transforms must satisfy
\begin{equation}
B_{\mathfrak{f}_4}(t) = B_{^L\mathfrak{f}_4}(t)\,,\qquad\qquad\quad B_{\mathfrak{g}_2}(t) = B_{^L\mathfrak{g}_2}(t)\,,
\end{equation}
so that the corresponding lattice sums \eqref{gsun2} reduce to
\begin{align}
\C_{G}(\tau)  &= \sum_{(m,n)\in \Z^2} \int_0^\infty \Big[ e^{- t \pi \frac{|m+n\tau|^2}{\tau_2}}+e^{- t \pi \frac{|m+\mathfrak{n}_{\mathfrak{g} } n\tau|^2}{\mathfrak{n}_{\mathfrak{g}}  \tau_2}} \Big]B_{\mathfrak{g}}(t) \,dt\,, \label{gFG}\\
&\notag = -2 b_{\mathfrak{g}}(0)+ \sum_{s=2}^\infty  b_{\mathfrak{g}}(s) \Big[\, {\rm E}(s;\tau) 
+{\rm E}(s;\mathfrak{n}_{\mathfrak{g} } \tau) \,\Big]\,,
\end{align}
which is manifestly invariant under $\Gamma_0(\mathfrak{n}_{\mathfrak{g}}=2)$ and $\Gamma_0(\mathfrak{n}_{\mathfrak{g}}=3)$ for $\mathfrak{g}=\mathfrak{f}_4$ and $\mathfrak{g}_2$ respectively since both ${\rm E}(s;\tau) $ and
 ${\rm E}(s;\mathfrak{n}_{\mathfrak{g} })$ separately are. 

The discussion is so far very similar to the classical cases $\mathfrak{so}_{2N+1}$ and $\mathfrak{sp}_{2N}$. 
An interesting novelty in the exceptional non-simply laced case, is that now $\hat S$ and $T$ generate an infinite group of symmetries forming a discrete subgroup of ${ SL}(2,\mathbb{R})$, denoted by $\Gamma_{(2,2\mathfrak{n}_\mathfrak{g},\infty)}$, which is not isomorphic to ${ SL}(2,\mathbb{Z})$ and provides a particular instance of  the more general notion of \textit{Hecke triangle group} or alternatively of \textit{Fricke group}.\footnote{Curiously in $\mathcal{N}=2$ SYM with $SU(N)$ gauge group and $2N$ fundamental flavours, other instances of Hecke triangle group were found to be realised as S-duality group acting on the effective coupling constant at special loci on the Coulomb branch \cite{Ashok:2016oyh}.}

Rather than discussing the most general definition, we can focus our attention on the triangle group $\Gamma_{(2,m,\infty)}$ with $m\in\mathbb{N}^{>1}$ relevant for the present discussion. 
The triangle group $\Gamma_{(2,m,\infty)}$ can be defined explicitly as the subgroup of $SL(2,\mathbb{R})$
\begin{equation}
\Gamma_{(2,m,\infty)} := \left\langle \mathcal{S},\,\mathcal{T},\,\mathcal{U}\,\,\Big\vert\,\,\mathcal{S}\mathcal{U}\mathcal{T}= \mathcal{S}^{2} = \mathcal{U}^{m} = - \mathbbm{1}_2 \right\rangle\,,\label{eq:triangle}
\end{equation}
where explicitly
\begin{equation}\label{eq:generators}
\mathcal{S}  = \left(\begin{matrix}
0& 1 \\ -1 & 0 \end{matrix}\right)\,,\qquad \mathcal{T} = \left(\begin{matrix}
1& 2\cos\big(\frac{\pi}{m}\big) \\ 0 & 1 \end{matrix}\right)\,,\qquad \mathcal{U}= \left(\begin{matrix}
0& 1 \\ -1 & 2\cos\big(\frac{\pi}{m}\big) \end{matrix}\right)\,.
\end{equation}
Note that $\Gamma_{(2,3,\infty)}\cong PSL(2,\Z)$ and $\mathcal{S}$, $\mathcal{U}(=\!\mathcal{S} \,(\mathcal{T}^{-1}))$ and $\mathcal{T}$ are precisely the order $2$, $3$ and $\infty$ generators of $ PSL(2,\Z)$.

The cases of interest are given by $m=2\mathfrak{n}_\mathfrak{g}$ which is either $4$ or $6$. The generators $\mathcal{S}$ and $\mathcal{T}$ defined in \eqref{eq:generators} can be understood from the GNO transformations \eqref{sduality} by rescaling the coupling constant $\tau\to \tilde{\tau}:= \sqrt{\mathfrak{n}_\mathfrak{g}} \tau$, yielding
\begin{align}
\hat{S}\cdot \tau & = -\frac{1}{\mathfrak{n}_\mathfrak{g} \tau} \qquad \Leftrightarrow \qquad\cS \cdot \tilde{\tau} = -\frac{1}{\tilde{\tau}}\,,\\
{T}\cdot \tau & = \tau+1\qquad\, \Leftrightarrow \qquad \cT \cdot \tilde{\tau} = \tilde{\tau} + 2\cos\big(\frac{\pi}{2\mathfrak{n}_\mathfrak{g}}\big)=\tilde{\tau} + \sqrt{\mathfrak{n}_\mathfrak{g}}\,.
\end{align}

Rather than redefining the coupling constant, we find it more convenient to describe the Fuchsian group $\Gamma_{(2,2 \mathfrak{n}_\mathfrak{g},\infty)}$ in terms of a conjugate \textit{Fricke group}.
The Fricke group of level $r\in \mathbb{N}^{>0}$ is a subgroup of $SL(2,\mathbb{R})$ generated by the congruence subgroup $\Gamma_0(r)$ and a matrix $S_r\in SL(2,\mathbb{R})$, called \textit{Fricke involution} and defined by
\begin{equation}
S_r := \left(\begin{matrix}
0& 1/ \sqrt{r} \\ -\sqrt{r} & 0 \end{matrix}\right)\,.
\end{equation}
Note that $S_r$ is indeed an involution,~i.e.~$S_r^2 = - \mathbbm{1}_2$, however, it is in general not an element of $SL(2,\mathbb{Z})$,~i.e.~$S_r \notin SL(2,\mathbb{Z})$ for $r\neq 1$.
For $r=2$ and $r=3$ the Fricke group of level $r$ is conjugate precisely to the triangle group $\Gamma_{(2,2r,\infty)}$. Furthermore we notice that in these particular cases the action of the Fricke involution, $S_r$, on $\tau$ exactly reproduces the GNO transformation $\hat{S}$ defined in \eqref{sduality}, i.e. $S_r\cdot\tau = -1 / (r\tau)$.

We arrive at the conclusion that the $\cN=4$ SYM integrated correlator \eqref{eq:firstcorr} with exceptional gauge groups $G=F_4$ and $G_2$ must be invariant under the GNO symmetry group,
\begin{equation}
\Gamma_{(2,2\mathfrak{n}_\mathfrak{g},\infty)} = \Gamma_0( \mathfrak{n}_\mathfrak{g}) \cup S_{\mathfrak{n}_\mathfrak{g}}\Gamma_0(\mathfrak{n}_\mathfrak{g})\label{eq:Fricke}\,.
\end{equation}
As already mentioned, our conjectural expression \eqref{gFG} is manifestly invariant under $\Gamma_0(\mathfrak{n}_{\mathfrak{g}})$. Furthermore, it is easy to see that under Fricke involution, $S_{\mathfrak{n}_\mathfrak{g}}$, the two lattice sums of exponential factors are exchanged,~i.e.
\begin{equation}
\Big[ \sum_{(m,n)\in \Z^2} e^{- t \pi \frac{|m+n\tau|^2}{\tau_2}}\Big]_{S_{\mathfrak{n}_\mathfrak{g}}} = \sum_{(\tilde{m},\tilde{n})\in \Z^2}  e^{- t \pi \frac{|\tilde{m}+ \tilde{n}\,\mathfrak{n}_{\mathfrak{g} }\tau|^2}{\mathfrak{n}_{\mathfrak{g}}\tau_2}}
\end{equation}
where the notation $[\cdots]_{\gamma}$ means that $\gamma$ acts on all occurrences of $\tau$ (and $\bar{\tau}$) inside the bracket.
Alternatively, from the formal sum over Eisenstein series we see that under Fricke involution the two Eisenstein series factors get again exchanged
\begin{equation}
 \Big[ {\rm E}(s;\tau) \Big]_{S_{\mathfrak{n}_\mathfrak{g}}} = {\rm E}(s;S_{\mathfrak{n}_\mathfrak{g}}\!\cdot\tau)={\rm E}\Big(s;-\frac{1}{\mathfrak{n}_{\mathfrak{g} } \tau}\Big)= {\rm E}(s;\mathfrak{n}_{\mathfrak{g} } \tau) \,.
\end{equation}

We conclude that our conjectural expression~\eqref{gFG}, for the $\cN=4$ SYM integrated correlator \eqref{eq:firstcorr} with exceptional gauge groups $F_4$ and $G_2$ has indeed the expected behaviour under generalised electro-magnetic duality~\eqref{sduality}.

Given that our expression \eqref{gFG} is transforming correctly under the expected GNO symmetries \eqref{sduality}, we can now try and see whether it is also compatible with our previous supersymmetric localisation calculations.
To this end, we repeat the same procedure outlined above. Firstly we use the supersymmetric localisation formula \eqref{eq:ZGpert}, or the universal expansion \eqref{pertexp}, to compute the perturbative expansion as $g_{_{Y\!M}}\to0$ of the exceptional integrated correlators.
Then, from the Fourier zero-mode expansion \eqref{eq:0mode}, we reconstruct $B_{\mathfrak{g}}(t)$. Once $B_{\mathfrak{g}}(t)$ is known, we use \eqref{gFG} to check whether our conjecture is valid non-perturbatively.

For $G=F_4$ we have $\mathfrak{n}_{\mathfrak{g}} \!=\!2$ and find
\begin{align}
\C_{F_4}(\tau) & =\sum_{(m,n)\in \Z^2} \int_0^\infty \Big[e^{- t \pi \frac{|m+n\tau|^2}{\tau_2}} +e^{- t \pi \frac{|m+2n\tau|^2}{2\tau_2}} \Big] B_{\mathfrak{f}_4}(t)\, dt\\
&\notag= 6 + \sum_{s=2}^\infty b_{F_4}(s) \Big[{\rm E}(s; \tau)+{\rm E}(s; 2\tau)\Big]\,,
\end{align}
with the $F_4$ Borel transform given by (remembering that $2h^\lor=18$ for $\mathfrak{f}_4$)
\begin{equation}
B_{\mathfrak{f}_4}(t) = \frac{ \cQ_{\mathfrak{f}_4}(t)}{(t+1)^{19}}\,,\label{eq:BF4}
\end{equation}
where we defined the palindromic polynomial
\begin{align}
\cQ_{\mathfrak{f}_4}(t) &:= 234 \big[ p_{\mathfrak{f}_4}(t) + t^{18} p_{\mathfrak{f}_4}(t^{-1})\big] = t^{18}\cQ_{\mathfrak{f}_4}(t^{-1}) \\
p_{\mathfrak{f}_4}(t) &\notag:=t-26 t^2+198 t^3-768 t^4+1923 t^5-4128 t^6+6438 t^7-8070t^8+4560 t^9\,.
\end{align}
From equation \eqref{eq:BF4}, it is a simple matter of calculations to show that $B_{\mathfrak{f}_4}(t)$ satisfies the identities \eqref{eq:Bident1}-\eqref{eq:Bident2}-\eqref{eq:Bint}.

Given $B_{\mathfrak{f}_4}(t) $, we can use \eqref{eqkinst} to compute the general $k$-instanton sector for the integrated correlator with gauge group $G=F_4$. In particular for $k=1$, the integral \eqref{eqkinst} can be easily evaluated to \eqref{eq:F4instEx}
thus identically matching the previous results obtained from the independent supersymmetric localisation calculation via the conjectural form \eqref{eq:ZnekS4} for exceptional Nekrasov partition function. 

Finally, for $G=G_2$ we have $\mathfrak{n}_{\mathfrak{g}} \!=\!3$ and find
\begin{align}
\C_{G_2}(\tau) & =\sum_{(m,n)\in \Z^2} \int_0^\infty \Big[e^{- t \pi \frac{|m+n\tau|^2}{\tau_2}} +e^{- t \pi \frac{|m+3n\tau|^2}{3\tau_2}} \Big] B_{\mathfrak{g}_2}(t) \,dt\\
&\notag= \frac{3}{2} + \sum_{s=2}^\infty b_{\mathfrak{g}_2}(s) \Big[{\rm E}(s; \tau)+{\rm E}(s; 3\tau)\Big]\,,
\end{align}
where the $G_2$ Borel transform is given by (remembering that $2h^\lor=8$ for $\mathfrak{g}_2$)
\begin{align}
B_{\mathfrak{g}_2}(t) &= \frac{\cQ_{\mathfrak{g}_2}(t)}{ (t+1)^9}\,.\label{eq:BG2}
\end{align}
having defined the palindromic polynomial 
\begin{align}
\cQ_{\mathfrak{g}_2}(t) &:= \frac{63}{2} \big[ p_{\mathfrak{g}_2}(t) + t^{8}  p_{\mathfrak{g}_2}(t^{-1})\big] = t^{8} \cQ_{\mathfrak{g}_2}(t^{-1})\,,\\
p_{\mathfrak{g}_2}(t)&\notag := t-11t^2+24 t^3 -12 t^4\,.
\end{align}
From \eqref{eq:BG2} we can then prove that $B_{\mathfrak{g}_2}(t) $ satisfies the identities \eqref{eq:Bident1}-\eqref{eq:Bident2}-\eqref{eq:Bint}.

Again, we can substitute the candidate expression for $B_{\mathfrak{g}_2}(t) $ in equation \eqref{eqkinst}  and predict the general $k$-instanton sector for the integrated correlator with gauge group $G_2$. For $k=1$, \eqref{eqkinst} reduces identically to the independent supersymmetric localisation calculation \eqref{eq:G2instEx} which we obtained from our conjectural expression \eqref{eq:ZnekS4} for exceptional Nekrasov partition function. 

Since both for $F_4$ and $G_2$ we have $\mathfrak{n}_\mathfrak{g}\neq 1$, we note that in our lattice sum calculation \eqref{eqkinst} only the first term contributes at the one-instanton level, while the second GNO-dual factor does not. As it is clear from  \eqref{eqkinst}, this second term will come into play at the two-instanton level for $F_4$ and at the three-instanton level for $G_2$. It would be extremely interesting to check that these terms do correctly reproduce the supersymmetric localisation results, unfortunately this would require having a formula for general Nekrasov partition function at higher instanton numbers, which is at the present time unknown. It should be possible to use Hilbert series methods \cite{Gaiotto:2012uq,Keller:2012da,Hanany:2012dm,Cremonesi:2014xha} to derive expressions for general Nekrasov partition function at higher instanton numbers. We are not aware of such results in $\cN=2^*$ SYM and arbitrary $\Omega$-deformation.

We stress however, that the Borel transforms $B_{\mathfrak{f}_4}(t)$ and $B_{\mathfrak{g}_2}(t)$ have been determined by matching \eqref{eq:0mode}, which does indeed depend crucially from the presence of the GNO-dual factor, with the perturbation expansion \eqref{pertexp}. Hence the non-perturbative results just presented do provide a completely independent check of our conjectural expressions, even if just at the one-instanton level.

\section{Laplace equations}
\label{sec:lapdiff}

In  \cite{Dorigoni:2021bvj,Dorigoni:2021guq}, the lattice sum formulation \eqref{gsun} was fundamental in establishing a striking property of the $SU(N)$ integrated correlator which is the existence of a Laplace equation that relates it to the $SU(N-1)$ and $SU(N+1)$ correlators. Similar Laplace difference equations \cite{Dorigoni:2022zcr} have been shown to hold as well for the integrated correlators with general classical gauge group. 

More in detail, we consider the action of the $SL(2,\Z)$-invariant hyperbolic laplacian, $\Delta_\tau=\tau_2^2 (\partial_{\tau_1}^2+\partial_{\tau_2}^2)$, on $\C_G(\tau)$ with $G=SU(N)$, $SO(n)$ (with $n=2N$ or $2N+1$) and $USp(2N)$.
Given the lattice sum representation \eqref{gsun}, it is easy to see that the action of $\Delta_\tau$ on $\C_G(\tau)$ can be translated into
\begin{equation}
\Delta_\tau \Big( e^{- t \pi \frac{|m+n\tau|^2}{\tau_2}} \Big)= t\, \partial_t^2 \Big(t e^{- t \pi \frac{|m+n\tau|^2}{\tau_2}}\Big) \,.
\end{equation} 
After an integration by parts we then obtain
\begin{equation}
\Delta_\tau \, \C_{G}(\tau) =  \sum_{(m,n)\in\mathbb{Z}^2}   \int_0^\infty  \left[ e^{ - t\,  Y_{mn}(\tau)} t\,\partial_t^2 \big(t B_{\mathfrak{g}}(t)\big) + e^{ - t\,  Y_{mn}(\mathfrak{n}_{\mathfrak{g}} \tau)}  t\,\partial_t^2 \big(t  B_{\!\,^L\mathfrak{g}}(t) \big) \right]  dt \,.
 \end{equation}

At this point, it was noted in \cite{Dorigoni:2021bvj,Dorigoni:2021guq,Dorigoni:2022zcr} that since the Borel transform $B_{\mathfrak{su}_N}(t)$ is related to Jacobi polynomials \eqref{eq:Bndef}, which are known to satisfy a three-term recursion relation, the action of the operator $t\partial_t^2 \big(t \cdot \big)$ on $B_{\mathfrak{su}_N}(t)$ had to close on the same space of Borel transforms, with a very similar argument holding for $B_{\mathfrak{so}_n}(t)$ (with $n=2N$ or $2N+1$) and $B_{\mathfrak{sp}_{2N}}(t)$.
This analysis led to the Laplace difference equations,
\begin{align}
&\Delta_\tau \, \C_{SU(N)}(\tau) \label{lapdiffSUN}  = 4c_{\mathfrak{su}_N}   \Big[\C_{SU(N+1)}(\tau)- 2\C_{SU(N)}(\tau)+\C_{SU(N-1)}(\tau)\Big]\\ 
&\nn\qquad\qquad\qquad \,\,\,\,+(N+1)\C_{SU(N-1)}(\tau) - (N-1) \C_{SU(N+1)}(\tau) \, ,\\
& \label{lapdiffSO}  \Delta_\tau \, \C_{SO(n)}(\tau)  =2 c_{\mathfrak{so}_n}  \Big[ \C_{SO(n+2)}(\tau) -2 \C_{SO(n)}(\tau) +\C_{SO(n-2)} (\tau)  \Big]  \\
&\nn  \qquad\qquad\qquad\,\,\,\,+ n\, \C_{SU(n-1)} (\tau) -(n-1) \C_{SU(n)} (\tau)  \, ,\\
&\label{lapdiffSP}\Delta_\tau \, \C_{USp(n)}(\tau) =2 c_{\mathfrak{sp}_n}  \Big[ \C_{USp(n+2)}(\tau) -2 \C_{USp(n)}(\tau)+\C_{USp(n-2)} (\tau) \Big] \\
&\nn \qquad\qquad\qquad \,\,\,\,-   n\, \C_{SU(n+1)} (2\tau)+ (n+1)\C_{SU(n)} (2\tau)   \, ,
\end{align}
with $c_{\mathfrak{g}} = \frac{{\rm dim}\mathfrak{g}}{4}$ the central charge of the theory.

These equations have powerful consequences. For a starter, given the initial condition $\C_{SU(1)}=0$, the first equation \eqref{lapdiffSUN} easily determines the correlator for gauge group $SU(N)$ in terms of the correlator for  gauge group $SU(2)$.  Furthermore it gives a very simple iterative procedure for determining terms in the  large-$N$ expansion of the correlator for gauge group $SU(N)$ both at the perturbative and non-perturbative \cite{Hatsuda:2022enx,Dorigoni:2022cua} level in $1/N$.   

 For the other classical Lie groups we first have the identities: 
 \begin{equation} \C_{USp(0)}(\tau)= \C_{SO(0)}(\tau) =\C_{SO(1)}(\tau)= \C_{SO(2)}(\tau) =0\,,
 \end{equation}
 which can be thought of as initial conditions.
 Then, combining such initial conditions with the fact\footnote{It should be emphasised that the initial conditions $\C_{SU(2)}(\tau)=\C_{SO(3)}(2\tau) = \C_{USp(2)}(\tau) $ are non-trivial properties. These identities have been checked at the perturbative and non-perturbative level in \cite{Dorigoni:2022zcr}.} that $\C_{SO(3)}(\tau)=\C_{SU(2)}(2\tau)$ we can use~\eqref{lapdiffSO}-\eqref{lapdiffSP} to determine $\C_{SO(2N)}(\tau)$, with $N\geq4$, purely as a linear combination of $\C_{SU(m)}(\tau)$ with $m=2,3,...,2N-2$, while $\C_{SO(2N+1)}(\tau)$ and  $\C_{USp(2N)}$, with $N\geq2$, again purely as linear combinations of $\C_{SU(m)}(\tau)$  and $\C_{SU(m)}(2\tau)$ with $m=2,3,..., 2N-1$.

Given our conjectural expression \eqref{gsun} for the exceptional integrated correlators written in terms of the explicit Borel transforms \eqref{eq:BE6}-\eqref{eq:BF4}-\eqref{eq:BG2}, we can repeat a very similar calculation to arrive at the exceptional Laplace equations (ordered in increasing level of ``complexity'')
\allowdisplaybreaks{
\begin{align}
& \Delta_\tau\, \C_{G_2}(\tau) = \frac{1}{5}\Big[132\,\C_{SU(2)}(\tau) +22\,\C_{SU(3)}(\tau) -123\, \C_{SU(4)}(\tau)+ 54\,\C_{SU(5)}(\tau) \Big]+ (\tau\to 3\tau)\,, \label{eq:LapExcG}\\
&\nn\Delta_\tau\, \C_{F_4}(\tau) =\\*
&\nn \frac{4}{1155} \Big[ 23031\, \C_{SU(2)}(\tau)-14139\,\C_{SU(3)}(\tau)-24219\,\C_{SU(4)}(\tau)+30969\, \C_{SU(5)}(\tau)-12753\,\C_{SU(6)}(\tau)\\*
&\qquad\,\,\,+28917\, \C_{SU(7)}(\tau)-28098 \,\C_{SU(8)}(\tau)-3808\,\C_{SU(9)}(\tau)+8064\,\C_{SU(10)}(\tau)\Big]+ (\tau\to 2\tau)\,, \label{eq:LapExcF}\\
&\nn\Delta_\tau \,\C_{E_6}(\tau) =\\*
&\nn\frac{1}{77} \Big[ 3600\, \C_{SU(2)}(\tau)-9222\, \C_{SU(3)}(\tau)+12501\, \C_{SU(4)}(\tau)-5382\, \C_{SU(5)}(\tau)+1116\, \C_{SU(6)}(\tau)\\*
&\nn\qquad -11268 \,\C_{SU(7)}(\tau)+10728\, \C_{SU(8)}(\tau)+3964\,\C_{SU(9)}(\tau)+252 \,\C_{SU(10)}(\tau) \\*
&\qquad-8766\,\C_{SU(11)}(\tau)+1221\,\C_{SU(12)}(\tau)+2178\, \C_{SU(13)}(\tau) \Big]\,.\label{eq:LapExcE}
\end{align}}

Surprisingly, using the Laplace-difference equation \eqref{lapdiffSUN} for the $SU(N)$ integrated correlator, we can actually solve the inhomogeneous Laplace equations \eqref{eq:LapExcG}-\eqref{eq:LapExcF}-\eqref{eq:LapExcE} and provide the equivalent algebraic identities,
\begin{align}
\C_{G_2}(\tau) &\label{eq:AlgExcG}= \frac{1}{10} \Big[-36 \, \C_{SU(2)}(\tau)+4\,  \C_{SU(3)}(\tau)+9 \, \C_{SU(4)}(\tau)\Big]+(\tau\to3\tau)\,,\\
\C_{F_4}(\tau)&\nn=-\frac{2}{1155} \Big[3033\,\C_{SU(2)}(\tau)-2322\,\C_{SU(3)}(\tau)-747\,\C_{SU(4)}(\tau)+1332\,\C_{SU(5)}(\tau)+261\,\C_{SU(6)}(\tau)\\
&\phantom{=}+306 \C_{SU(7)}(\tau)-504\, \C_{SU(8)}(\tau)-224\, \C_{SU(9)}(\tau)\Big]+(\tau\to2\tau)\,,\label{eq:AlgExcF}\\
\C_{E_6}(\tau)&\nn=\frac{1}{154} \Big[-336\,  \C_{SU(2)}(\tau)+432  \,\C_{SU(3)}(\tau)-543 \, \C_{SU(4)}(\tau)+204\,  \C_{SU(5)}(\tau)+132\,  \C_{SU(6)}(\tau)\\
&\label{eq:AlgExcE}\!\!\!\!\!\!\!\!\!\!\!\!\!\!+132 \, \C_{SU(7)}(\tau)-192\,  \C_{SU(8)}(\tau)-92\,  \C_{SU(9)}(\tau)+36 \, \C_{SU(10)}(\tau)+108\,  \C_{SU(11)}(\tau)+33\,  \C_{SU(12)}(\tau)\Big]\,.
\end{align}

Given the algebraic relations \eqref{eq:AlgExcG}-\eqref{eq:AlgExcF}-\eqref{eq:AlgExcE}, it should be possible to rewrite the exceptional Laplace equations \eqref{eq:LapExcG}-\eqref{eq:LapExcF}-\eqref{eq:LapExcE} in a form more closely resembling \eqref{lapdiffSUN}-\eqref{lapdiffSP}-\eqref{lapdiffSO} where, even for the exceptional cases $G=G_2$, $F_4$ and $E_6$, the source terms in \eqref{eq:LapExcG}-\eqref{eq:LapExcF}-\eqref{eq:LapExcE} are written in terms of a suitable multiple of $c_{\mathfrak{g}}\, \C_G(\tau)$, with $c_{\mathfrak{g}}$ the associated central charge, plus ``natural'' linear combinations of $\C_{SU(m)}$ correlators for certain ranges of $m$.

We believe there should be a universal way, in the sense of Vogel, to rewrite all Laplace difference equations \eqref{lapdiffSUN}-\eqref{lapdiffSP}-\eqref{lapdiffSO}-\eqref{eq:LapExcG}-\eqref{eq:LapExcF}-\eqref{eq:LapExcE} in terms of a single universal Laplace equation expressed in terms of Vogel parameters, very much in the same spirit as our discussion in section \ref{pertexpn}. The remaining missing correlators $\C_{E_7}$ and $\C_{E_8}$ would arise naturally as special points of this universal Laplace equation. We hope to solve this problem in future work.

\section{Discussion }
\label{sec:discussion}

In this work we have extended the analysis initiated in \cite{Dorigoni:2021bvj,Dorigoni:2021guq, Dorigoni:2022zcr}  and we have proposed a lattice sum representation for the integrated correlator, $\mathcal{C}_G(\tau)$, of four superconformal primary in the stress tensor multiplet of $\cN=4$ SYM  and valid for arbitrary simple gauge group $G$. These integrated correlators are determined via supersymmetric localisation and manifest a beautiful set of properties which reflects the constraints imposed by GNO duality. In particular, we have discovered a new set of inhomogeneous Laplace equations satisfied by the integrated correlator  $\mathcal{C}_G(\tau)$ with exceptional gauge groups $G=G_2,$ $F_4,$ and $E_6$.

To support our claims we have performed perturbative and non-perturbative checks.
Having defined a suitable 't Hooft-like coupling, $a_{\mathfrak{g}}$, we have shown that the perturbation expansion of $\mathcal{C}_G(\tau)$ is universal in the sense of Vogel. We have proposed a single unifying perturbative expansion which does not require to specify a particular gauge group. By evaluating this universal expression at special points on the Vogel plane we have managed to obtain the weak-coupling expansions of the integrated correlator for all simple gauge groups.

At the non-perturbative level, we have relied crucially on work carried out in \cite{Billo:2015pjb,Billo:2015jyt,Billo:2016zbf} to propose a candidate expression for the one-instanton Nekrasov partition function for $\cN=2^*$ SYM in the presence of an $\Omega$-deformation background and valid for a generic simple gauge group. This conjectural general Nekrasov partition function does reproduce the correct known results in absence of $\Omega$-deformation and in the limit of an infinitely massive hypermultiplet. Furthermore we have checked the consistency between the proposed lattice sum representation for $\mathcal{C}_G(\tau)$ and the one-instanton contributions which can be obtained via supersymmetric localisation from the proposed Nekrasov partition function.

We conclude by mentioning a couple of directions in need of further investigations. 
Firstly, it would be interesting to understand whether Vogel universality for the integrated correlator extends to the non-perturbative level. From the work of Vogel which remains unfinished and unpublished, it is not quite clear which Lie algebraic quantities enjoy universal expressions. In particular, it would be remarkable to understand whether all different instances of inhomogeneous Laplace equations, found for the various specific gauge groups,~$G$, can all be encoded universally in a single inhomogeneous Laplace equation for the integrated correlator corresponding to the universal Lie algebra.

Secondly, focus of the present work is the integrated correlator $\mathcal{C}_G(\tau)$ which is directly related to a four-point function of local operators and, for this reason, insensitive to global properties of the gauge group $G$. As a consequence of this fact, GNO duality can be recast at the level of Lie algebras thus leading to interesting, yet simpler transformation properties of $\mathcal{C}_G(\tau)$ under S-duality. Extended operators in $\cN=4$ SYM are extremely important physical quantities which depend crucially on global properties of $G$ and $^LG$. 
Very recently in \cite{Pufu:2023vwo} the authors studied an integrated correlator of two superconformal primaries in the stress tensor multiplet in the presence of a half-BPS Wilson line defect in $\cN=4$ SYM with gauge group $SU(N)$ at large-$N$. Similar results are not known for general gauge groups and do most definitely deserve further studies for general gauge group $G$ since under S-duality a Wilson line in $\cN=4$ SYM with gauge group $G$ is mapped into a 't Hooft line defect in the $^LG$ theory.

 \section*{Acknowledgements}
We would like to particularly thank Axel Kleinschmidt and Boris Pioline for bringing to our attention the potential use of Vogel universality for the study of integrated correlators with exceptional gauge groups. We also thank Stefano Cremonesi, Marialuisa Frau, Alberto Lerda, Ruben Mkrtchyan and Alessandro Pini for useful discussions and Marialuisa Frau, Axel Kleinschmidt, Alberto Lerda and Congkao Wen for comments on the draft.
 DD would like to thank the Albert Einstein Institute, Golm, for the hospitality during the final stages of this project.
PV would like to thank the Department of Mathematical Sciences of Durham University for the hospitality and support within the scientific visits scheme organised by the International Network on Quantum Fields and Strings (IRN:QFS). PV is partially supported by the MUR PRIN contract 2020KR4KN2 ``String Theory as a bridge between Gauge Theories and Quantum Gravity'' and by the INFN project ST\&FI ``String Theory \& Fundamental Interactions''.
 

\appendix

\section{Root systems for simple Lie algebras}
\label{app:Lie}

In this appendix we provide a list of root systems for all simple Lie algebras, $\mathfrak{g}$, considered in this paper. 
We will write all roots in terms of the $\mathbb{R}^r$ orthonormal basis $\{ {\bf e}_1,... {\bf e}_r\}$ with $r=\mbox{rank}(\mathfrak{g})$.
We denote by $\Delta$ the complete root system, while $\Delta^+$ denotes only the positive roots which we further divide into short, $\Delta_S^{+}$ and long roots, $\Delta_L^+$.

\begin{itemize}
\item $A_n = \mathfrak{su}_{n+1}$. The positive roots are
\begin{equation}
\Delta^+  = \{{ \bf e}_i -{\bf e}_j\,:\, 1\leq i < j \leq n+1\}\,.
\end{equation}
Since $A_n$ is simply laced, all of its roots are by convention long roots hence the ratio between the length square of long and short roots is
\begin{equation}
\mathfrak{n}_{\mathfrak{su}_{n}} = 1\,.
\end{equation}
\item $B_n = \mathfrak{so}_{2n+1}$. The short positive roots and the long positive roots are given by
\begin{equation}
\Delta_S^+ = \{\sqrt{2} {\bf e}_i\,:\, 1\leq i \leq n\} \,,\qquad \Delta_{L}^+ = \{\sqrt{2}\,{\bf e}_i \pm \sqrt{2}\,{\bf e}_j\,:\, 1\leq i < j \leq n\}\,.
\end{equation}
Note that the ratio between the length square of long and short roots is
\begin{equation}
\mathfrak{n}_{\mathfrak{so}_{2n+1}} = \frac{|\alpha_{\rm{long}}|^2}{|\alpha_{\rm short}|^2} = 2\,,
\end{equation}
where $\alpha_{\rm{long}} \in \Delta_L^+$ and $\alpha_{\rm{short}}\in \Delta_S^+$.

\item $C_n = \mathfrak{sp}_{2n}$. The short positive roots and the long positive roots are given by
\begin{equation}
\Delta_S^+ = \{{\bf e}_i \pm {\bf e}_j\,:\, 1\leq i < j \leq n\} \,,\qquad \Delta_{L}^+ =\{2\, {\bf e}_i\,:\, 1\leq i \leq n\} \,.
\end{equation}
Note that the ratio between the length square of long and short roots is
\begin{equation}
\mathfrak{n}_{\mathfrak{sp}_{2n}} = \frac{|\alpha_{\rm{long}}|^2}{|\alpha_{\rm short}|^2} = 2\,,
\end{equation}
where $\alpha_{\rm{long}} \in \Delta_L^+$ and $\alpha_{\rm{short}}\in \Delta_S^+$.

\item $D_n = \mathfrak{so}_{2n}$. The positive roots are
\begin{equation}
\Delta^+  = \{{ \bf e}_i \pm {\bf e}_j\,:\, 1\leq i < j \leq n\}\,.
\end{equation}
Since $D_n$ is simply laced, all of its roots are by convention long roots hence the ratio between the length square of long and short roots is
\begin{equation}
\mathfrak{n}_{\mathfrak{so}_{2n}} = 1\,.
\end{equation}

\item $E_6$. The $36$ positive roots of $E_6$ are
\begin{equation}
\Delta^+ = \{{ \bf e}_i \pm {\bf e}_j\,:\, 1\leq i < j \leq 5\} \cup \Big\{\frac{1}{2}(\pm {\bf e}_1 \pm {\bf e}_2 \pm {\bf e}_3 \pm {\bf e}_4 \pm {\bf e}_5 +\sqrt{3}\, {\bf e}_6)\Big\} _{\mbox{\scriptsize{\# minus signs even}}}\,.
\end{equation}

\item $E_7$. The $63$ positive roots of $E_7$ are
\begin{equation}
\Delta^+ {=} \{{ \bf e}_i \pm {\bf e}_j: 1\leq i < j \leq 6\} \cup \{\sqrt{2}\, {\bf e}_7\}\cup \Big\{\frac{1}{2}(\pm {\bf e}_1 \pm {\bf e}_2 \pm {\bf e}_3 \pm {\bf e}_4 \pm {\bf e}_5  \pm {\bf e}_6+\sqrt{2} \,{\bf e}_7)\Big\} _{\mbox{\scriptsize{\# minus signs odd}}}\,.\label{eq:e7roots}
\end{equation}

\item $E_8$. The $120$ positive roots of $E_8$ are
\begin{equation}
\Delta^+ = \{{ \bf e}_i \pm {\bf e}_j\,:\, 1\leq i < j \leq 8\} \cup \Big\{\frac{1}{2}(\pm {\bf e}_1 \pm {\bf e}_2 \pm {\bf e}_3 \pm {\bf e}_4 \pm {\bf e}_5 \pm {\bf e}_6\pm {\bf e}_7 + {\bf e}_8)\Big\} _{\mbox{\scriptsize{\# minus signs even}}}\,.\label{eq:e8roots}
\end{equation}

Since $E_{6,7,8}$ are all simply laced, all of their roots are by convention long roots hence the ratio between the length square of long and short roots is
\begin{equation}
\mathfrak{n}_{\mathfrak{e}_{{\rm r}}} = 1\,,\qquad\qquad {\rm r}\in \{6,7,8\}\,.
\end{equation}

\item $F_4$. The short positive roots and the long positive roots are given by
\begin{equation}
\Delta_S^+ = \{\sqrt{2} \,{\bf e}_i \,:\, 1 \leq i \leq 4 \}\cup\Big\{\frac{1}{\sqrt{2}}({\bf e}_1 \pm {\bf e}_2 \pm {\bf e}_3 \pm {\bf e}_4 )\Big\} \,,\qquad \Delta_{L}^+ =\{ \sqrt{2}\,( {\bf e}_i \pm {\bf e}_j) \,:\, 1\leq i<j \leq 4\} \,.
\end{equation}
Note that the ratio between the length square of long and short roots is
\begin{equation}
\mathfrak{n}_{\mathfrak{f}_{4}} = \frac{|\alpha_{\rm{long}}|^2}{|\alpha_{\rm short}|^2} = 2\,,
\end{equation}
where $\alpha_{\rm{long}} \in \Delta_L^+$ and $\alpha_{\rm{short}}\in \Delta_S^+$.

\item $G_2$. The short positive roots and the long positive roots are given by
\begin{equation}
\Delta_S^+ = \Big\{\sqrt{2} \,{\bf e}_1 \,,\, \pm\frac{1}{\sqrt{2}} {\bf e}_1 +\sqrt{\frac{3}{2}} {\bf e}_2\Big\} \,,\qquad \Delta_{L}^+ =\Big\{\pm\frac{3}{\sqrt{2}} {\bf e}_1 + \sqrt{\frac{3}{2}} {\bf e}_2 \,,\, \sqrt{6} \,{\bf e}_2\Big\} \,.
\end{equation}
Note that the ratio between the length square of long and short roots is
\begin{equation}
\mathfrak{n}_{\mathfrak{g}_{2}} = \frac{|\alpha_{\rm{long}}|^2}{|\alpha_{\rm short}|^2} = 3\,,
\end{equation}
where $\alpha_{\rm{long}} \in \Delta_L^+$ and $\alpha_{\rm{short}}\in \Delta_S^+$.

\end{itemize}

\section{Universal perturbative expansion at higher orders}
\label{app:UniPert}

In this appendix we present higher order terms in the perturbative weak-coupling expansion \eqref{pertexp} of the integrated correlator  expressed in terms of Vogel parameters $(\alpha,\beta,\gamma)$ parametrising the general universal Lie algebra, $\mathfrak{g}$.
We refer to Table \ref{tab:Lie}-(b) for the specific points in Vogel plane which corresponds to the simple Lie algebras of interest for the present work.

Firstly as in the main text, we choose variables 
$$\sigma_i := \alpha^i+\beta^i+\gamma^i\,,\qquad i=1,2,3\,,$$
 as a basis for all symmetric polynomials in $(\alpha,\beta,\gamma)$. Given that Vogel plane is $\mathbb{P}^2/\mathfrak{S}_3$, we note that under rescaling $(\alpha,\beta,\gamma)\to (k\alpha,k\beta,k\gamma)$ we have $\sigma_i\to k^i \sigma_i$. We recall that $P_k(\mathfrak{g})= 0$ for $k=1,2,3$, while for higher order universal terms we find,
\allowdisplaybreaks{
\begin{align*}
&P_4(\mathfrak{g})=-\frac{2 \sigma _1^3-3 \sigma _2 \sigma _1+\sigma _3}{84 \,\sigma _1^3}\,,\\
&P_5(\mathfrak{g})=-\frac{2 \sigma _1^3-3 \sigma _2 \sigma _1+\sigma _3}{24 \,\sigma _1^3}\,,\\
&P_6(\mathfrak{g})=-\frac{96 \sigma _1^5-145 \sigma _2 \sigma _1^3+51 \sigma _3 \sigma _1^2-3 \sigma _2^2 \sigma _1+\sigma _2 \sigma _3}{528 \sigma _1^5}\,,\\
&P_7(\mathfrak{g})=-\frac{10776 \sigma _1^6-16201 \sigma _2 \sigma _1^4+6171 \sigma _3 \sigma _1^3-1245 \sigma _2^2 \sigma _1^2+541 \sigma _2 \sigma _3 \sigma _1-42 \sigma _3^2}{34320\, \sigma _1^6}\,,\\
&P_8(\mathfrak{g})=-\frac{6408 \sigma _1^7-9337 \sigma _2 \sigma _1^5+3891 \sigma _3 \sigma _1^4-1776 \sigma _2^2 \sigma _1^3+940 \sigma _2 \sigma _3 \sigma _1^2-27 \sigma _2^3 \sigma _1-108 \sigma _3^2 \sigma _1+9 \sigma _2^2 \sigma _3}{13728 \,\sigma _1^7}\,,\\
&P_9(\mathfrak{g})=-\frac{52776 \sigma _1^8-71443 \sigma _2 \sigma _1^6+32673 \sigma _3 \sigma _1^5-28236 \sigma _2^2 \sigma _1^4+17614 \sigma _2 \sigma _3 \sigma _1^3-1593 \sigma _2^3 \sigma _1^2}{84864\, \sigma _1^8}\\
&\qquad\qquad\!\!\!+\frac{2502 \sigma _3^2 \sigma _1^2-801 \sigma _2^2 \sigma _3 \sigma _1+90 \sigma _2 \sigma _3^2}{84864\, \sigma _1^8}\,,\\
&P_{10}(\mathfrak{g})=-\frac{25738744 \sigma _1^9-30426993 \sigma _2 \sigma _1^7+15286267 \sigma _3 \sigma _1^6-22819188 \sigma _2^2 \sigma _1^5+16446360 \sigma _2 \sigma _3 \sigma _1^4}{33860736\, \sigma _1^9}\\*
&\qquad\qquad\!+\frac{3206637 \sigma _2^3 \sigma _1^3+2745788 \sigma _3^2 \sigma _1^3-2126523 \sigma _2^2 \sigma _3 \sigma _1^2+39690 \sigma _2^4 \sigma _1+388848 \sigma _2 \sigma _3^2 \sigma _1}{33860736 \,\sigma _1^9}\\*
&\qquad\qquad\! -\frac{16020 \sigma _3^3+13230 \sigma _2^3 \sigma _3}{33860736 \,\sigma _1^9}\,,\\
&P_{11}(\mathfrak{g})=-\frac{2257936 \sigma _1^{10}-2130762 \sigma _2 \sigma _1^8+1178230 \sigma _3 \sigma _1^7-2898399 \sigma _2^2 \sigma _1^6+2383335 \sigma _2 \sigma _3 \sigma _1^5-858096 \sigma _2^3 \sigma _1^4}{2604672\, \sigma _1^{10}}\\*
&\qquad\qquad \! +\frac{457958 \sigma _3^2 \sigma _1^4-704568 \sigma _2^2 \sigma _3 \sigma _1^3+39447 \sigma _2^4 \sigma _1^2+169872 \sigma _2 \sigma _3^2 \sigma _1^2-11376 \sigma _3^3 \sigma _1}{2604672\, \sigma _1^{10}}\\*
&\qquad\qquad\! -\frac{22059 \sigma _2^3 \sigma _3 \sigma _1-2970 \sigma _2^2 \sigma _3^2}{2604672\, \sigma _1^{10}}\,,\\
&P_{12}(\mathfrak{g})=-\frac{160273024 \sigma _1^{11}-105818280 \sigma _2 \sigma _1^9+64726672 \sigma _3 \sigma _1^8-258101472 \sigma _2^2 \sigma _1^7+240613575 \sigma _2 \sigma _3 \sigma _1^6}{171164160\, \sigma _1^{11}}\\*
&\qquad\qquad\! +\frac{148623777 \sigma _2^3 \sigma _1^5+52687325 \sigma _3^2 \sigma _1^5-145879266 \sigma _2^2 \sigma _3 \sigma _1^4+17134335 \sigma _2^4 \sigma _1^3}{171164160\, \sigma _1^{11}}\\*
&\qquad\qquad\! + \frac{ 43221915 \sigma _2 \sigma _3^2 \sigma _1^3-3783474 \sigma _3^3 \sigma _1^2-13330467 \sigma _2^3 \sigma _3 \sigma _1^2+182250 \sigma _2^5 \sigma _1+}{171164160\, \sigma _1^{11}}\\*
&\qquad\qquad\!+\frac{3103974 \sigma _2^2 \sigma _3^2 \sigma _1-206100 \sigma _2 \sigma _3^3-60750 \sigma _2^4 \sigma _3}{171164160\, \sigma _1^{11}}\,.
\end{align*}}

Secondly, we present the very same higher-order terms expressed in a different basis of symmetric polynomials in three variables, namely we define 
\begin{equation}
t:=\alpha+\beta+\gamma\,,\qquad s:=\alpha\beta+\alpha\gamma+\beta\gamma\,,\qquad p:=\alpha\beta\gamma\,,
\end{equation} although $t=\sigma_1$ it is conventional to keep $t$ as notation.
We note again that under rescaling $t\to k^1 t,$ $s\to k^2 s$ and $p\to k^3 p$.
In this alternative basis we have
\allowdisplaybreaks{\begin{align*}
&P_4(\mathfrak{g})=-\frac{p+s t}{28\, t^3}\,,\\
&P_5(\mathfrak{g})=-\frac{p+s t}{8\, t^3}\,,\\
&P_6(\mathfrak{g})=\frac{p s-26 p t^2+s^2 t-24 s t^3}{88\, t^5}\,,\\
&P_7(\mathfrak{g})=\frac{63 p^2+415 p s t-3314 p t^3+352 s^2 t^2-2736 s t^4}{5720\, t^6}\,,\\
&P_8(\mathfrak{g})=\frac{81 p^2 t-9 p s^2+317 p s t^2-1156 p t^4-9 s^3 t+218 s^2 t^3-846 s t^5}{1144\, t^7}\,,\\
&P_9(\mathfrak{g})=-\frac{135 p^2 s-1944 p^2 t^2+531 p s^2 t-5630 p s t^3+11476 p t^5+396 s^3 t^2-3164 s^2 t^4+7452 s t^6}{7072\, t^8}\,,\\
&P_{10}(\mathfrak{g})=-\frac{12015 p^3+158379 p^2 s t-771644 p^2 t^3-8820 p s^3+369268 p s^2 t^2-1783612 p s t^4+2304264 p t^6}{940576\, t^9}\\*
&\qquad\qquad\! +\frac{8820 s^4 t-205264 s^3 t^3+834858 s^2 t^5-1334280 s t^7}{940576\, t^9}\,,\\
&P_{11}(\mathfrak{g})=-\frac{8532 p^3 t-2970 p^2 s^2+62310 p^2 s t^2-149168 p^2 t^4-8766 p s^3 t+104719 p s^2 t^3-286168 p s t^5}{72352\, t^{10}}\\*
&\qquad \qquad \!+\frac{-255060 p t^7+5796 s^4 t^2-45289 s^3 t^4+113730 s^2 t^6-132540 s t^8}{72352\, t^{10}}\,,\\
&P_{12}(\mathfrak{g})=\frac{618300 p^3 s-5984361 p^3 t^2+4353048 p^2 s^2 t-30858480 p^2 s t^3+43522246 p^2 t^5-162000 p s^4}{9509120\, t^{11}}\\*
&\qquad\qquad\!+\frac{-7536960 p s^3 t^2+40348103 p s^2 t^4-71077202 p s t^6+46425504 p t^8+162000 s^5 t-3478212 s^4 t^3}{9509120\, t^{11}}\\*
&\qquad\qquad \! +\frac{14102456 s^3 t^5-24393912 s^2 t^7+21792096 s t^9}{9509120\, t^{11}}\,.
\end{align*}}

\bibliographystyle{ssg}
\bibliography{cites}
	
\end{document}